\documentclass[twocolumn,english,american]{iopart}
\usepackage[T1]{fontenc}
\usepackage[latin9]{inputenc}
\usepackage{geometry}
\usepackage{amssymb}
\usepackage{graphicx}

\makeatletter
\usepackage{iopams}
\usepackage{setstack}

\usepackage{fancyhdr}
\usepackage[numbers,sort&compress]{natbib}

\makeatother

\usepackage{babel}
\begin{document}

\title{Spin, orbital and topological order in models of strongly correlated
electrons}

\author{Wojciech Brzezicki}

\address{International Research Centre MagTop at Institute of Physics, Polish
Academy of Sciences, Aleja Lotnik\'ow 32/46, PL-02668 Warsaw, Poland}
\begin{abstract}
Different types of order are discussed in the context of strongly
correlated transition metal oxides, involving pure compounds and $3d^{3}-4d^{4}$
and $3d^{2}-4d^{4}$ hybrids. Apart from standard, long-range spin
and orbital orders we observe also exotic non-colinear spin patterns.
Such patters can arise in presence of atomic spin-orbit coupling,
which is a typical case, or due to spin-orbital entanglement at the
bonds in its absence, being much less trivial. Within a special interacting
one-dimensional spin-orbital model it is also possible to find a rigorous
topological magnetic order in a gapless phase that goes beyond any
classification tables of topological states of matter. This is an
exotic example of a strongly correlated topological state. Finally,
in the less correlated limit of $4d^{4}$ oxides, when orbital selective
Mott localization can occur it is possible to stabilize by a $3d^{3}$
doping one-dimensional zigzag antiferromagnetic phases. Such phases
have exhibit nonsymmorphic spatial symmetries that can lead to various
topological phenomena, like single and mutliple Dirac points that
can turn into nodal rings or multiple topological charges protecting
single Dirac points. Finally, by creating a one-dimensional $3d^{2}-4d^{4}$
hybrid system that involves orbital pairing terms, it is possible
to obtain an insulating spin-orbital model where the orbital part
after fermionization maps to a non-uniform Kitaev model. Such model
is proved to have topological phases in a wide parameter range even
in the case of completely disordered $3d^{2}$ impurities. What more,
it exhibits hidden Lorentz-like symmetries of the topological phase,
that live in the parameters space of the model. 
\end{abstract}
\maketitle

\section{Introduction}

\begin{figure*}[t]
\begin{center} \includegraphics[width=1\textwidth]{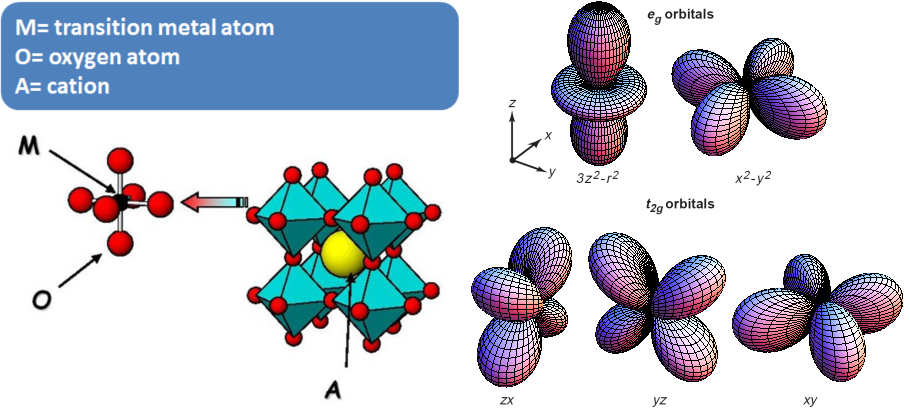}\end{center}\caption{Perovskite structure of transition metal oxides (left); metal ions
are surrounded by oxygen octahedra and form a cubic lattice. View
of orbitals $d$ in a cubic surrounding (right); $e_{g}$ orbitals
doublet and $t_{2g}$ orbitals triplet.\label{fig:1}}
\end{figure*}

The effects of strong correlations are typically observed in transition
metal oxides (TMOs) of perovskite structure, see Fig. \ref{fig:1},
where transition metal ions are enclosed in oxygen octahedra and form
a cubic lattice \cite{Ole05}. Transition metal ions are characterized
by not fully filled $d$ shell what determines their chemical properties.
Orbitals $d$ have orbital quantum number $l=2$ and thus five possible
states with magnetic quantum number $m$. These five states in case
of isolated atom are degenerate but in case of an atom in cubic environment
they get split into three $t_{2g}$ states having lower energy and
two $e_{g}$ states with higher energy, see Fig. \ref{fig:1}. Hence,
there is still orbital degeneracy left in a cubic crystal. $t_{2g}$
states are typically labeled as; $|zx\rangle$, $|yz\rangle$ and
$|xy\rangle$, while $e_{g}$ states are denoted as; $|3z^{2}-r^{2}\rangle$
and $|x^{2}-y^{2}\rangle$. A commonly used convention for $t_{2g}$
orbitals is; $|a\rangle\equiv|zx\rangle$, $|b\rangle\equiv|yz\rangle$
and $|c\rangle\equiv|xy\rangle$ \textendash{} its meaning will become
clear later.

Electrons in TMOs can move mainly by hopping from metal to oxygen
ions and vice versa what follows from the overlap of their atomic
wave functions. These electrons are subject to Coulomb repulsion which
is assumed to be purely local due to screening, i.e., electrons interact
only when they are at the same atom. In many cases, that depend on
the band structure \cite{Zaa85}, interactions at the oxygen ions
are not relevant for the ground state and the system can be described
only by a lattice of transition metal ions. The hopping between them
is possible due to hybridization with oxygen atoms. Electrons that
are at the same atom or lattice site can interact being in the same
orbital state and different spin states, then we talk about Hubbard
interaction $U$, or can interact being in different orbital states
and then the Coulomb interaction depends on spin configuration of
electrons via Hund's exchange $J_{H}$ and is lowered by $J_{H}$
for parallel spins \cite{Kug73}. A generic interaction Hamiltonian
for transition metal ions has a form of,

\begin{eqnarray}
\fl H_{int} & \!\!=\!\! & \sum_{i}\left\{ U\sum_{\mu}n_{i\mu\uparrow}n_{i\mu\downarrow}\!+\!\!J_{H}\sum_{\mu\not=\nu}d_{i\mu\uparrow}^{\dagger}d_{i\mu\downarrow}^{\dagger}d_{i\nu\downarrow}d_{i\nu\uparrow}\right.\nonumber \\
\fl & \!\!+\!\! & \left.\left(\!U\!-\!\frac{5}{2}J_{H}\!\right)\!\!\sum_{{\mu<\nu\atop \sigma\sigma'}}n_{i\mu\sigma}n_{i\nu\sigma'}\!-\!2J_{H}\!\sum_{\mu<\nu}{\bf S}_{i\mu}\!\cdot\!{\bf S}_{i\nu}\right\} \label{eq:Hint}
\end{eqnarray}
where fermion operator $d_{i\mu\sigma}^{\dagger}$ creates an electron
with spin $\sigma$ in orbital $\mu$ and ${\bf S}_{i}$ is a spin
operator defined as ${\bf S}_{i}\!=\!\frac{1}{2}\!\sum_{\mu,\alpha,\beta}\!d_{i\mu\alpha}^{\dagger}{\bf \sigma}_{\!\alpha,\beta}d_{i\mu\beta}$.
Values of $U$ i $J_{H}$ differ for different transition metals,
but $J_{H}$ coupling always favors state with maximal spin $S$,
which is the Hund's rule. The kinetic Hamiltonian can be written as,

\begin{equation}
\fl H_{t}=-\sum_{{\left\langle i,j\right\rangle \parallel\gamma\atop \sigma}}t_{\mu\nu}^{\gamma}\left(d_{i\mu\sigma}^{\dagger}d_{j\nu\sigma}+d_{j\nu\sigma}^{\dagger}d_{i\mu\sigma}\right),\label{eq:Ht}
\end{equation}
where the sum is over the nearest neighbors (NNs) $\langle i,j\rangle$
and hopping amplitudes $t_{\mu\nu}^{\gamma}$ depends on the bond's
direction $\hat{\gamma}$ and a pair of orbitals between which the
hopping takes place. This follows from the fact that orbitals live
in real space (unlike spins) and their overlap depend on $\hat{\gamma}$.
For example, in case of $t_{2g}$ orbitals for a bond in $\hat{a}$
direction the only non-vanishing hopping amplitudes are $t_{bb}=t_{cc}=t$
and, in general, for any direction $\hat{\gamma}$ there is no hopping
between two orbitals $\gamma$. This is a consequence of their symmetry
and symmetry of a cubic lattice, which can be changed for instance
by a lattice distortions, and also of participation of oxygen orbitals
$2p_{\pi}$ in hopping processes $d-d$.

In case when interaction between electrons are strong and there is
no doping changing the number of electrons at the lattice sites, a
Mott localization can occur \cite{Mot68} being a metal-insulator
transition only due to electron-electron correlations, with crucial
role played by antiferromagnetic spin exchange $J\propto t^{2}/U$
\cite{Wys17}. Such an insulator can be described by an effective
model where charge degrees of freedom are absent \cite{Ole83,Kha05}.
The charge motion is possible only within virtual superexchange processes,
when an electron hops from on metal ion to another one (via oxygen)
and hops back. An effective superexchange Hamiltonian can be derived
by a perturbative expansion where a perturbation is the kinetic Hamiltonian
$H_{t}$ and unperturbed states are the spin and orbital degenerate
eigenstates of $H_{int}$. As a result we obtain a model of interacting
spin and orbitals known as spin-orbital superexchange model, with
the simplest realization being a Kugel-Khomskii model \cite{Kug73}
of the generic form of, 

\begin{equation}
\fl{\cal H}=\sum_{\left\langle i,j\right\rangle }\left\{ J_{i,j}\left(\vec{T}_{i},\vec{T}_{j}\right)\vec{S}_{i}\cdot\vec{S}_{j}+K_{i,j}\left(\vec{T}_{i},\vec{T}_{j}\right)\right\} ,\label{eq:sex}
\end{equation}
where spin $\vec{S}_{i}$ stands for total spin at the transition
metal ion following from the Hund's rule, $\vec{T_{i}}$ are pseudospin
operators describing orbital degrees of freedom and $J_{i,j}$, $K_{i,j}$
are some bilinear functions of these operators depending on the bond's
$\langle i,j\rangle$ direction. For metals with high main quantum
number, i.e., $4d$ or $5d$ ones, one should also include a finite
spin-orbit coupling (SOC) $\lambda$. Such a coupling can be added
to the superexchange Hamiltonian (\ref{eq:sex}) as a $\lambda\sum_{i}\vec{S}_{i}\cdot\vec{L}_{i}$,
where $\vec{L}_{i}$ is an angular momentum operator that can be expressed
in terms of components of $\vec{T_{i}}$ \cite{Brz15X}, provided
that $\lambda\ll J_{H}$ \textendash{} otherwise the high spin description
following from $H_{int}$ is not correct and one has to start from
local spin-orbital entangled eigenstates determined by $\lambda$
\cite{Jac09}.

Equation (\ref{eq:sex}) defines a highly non-trivial quantum many-body
problem whose complexity grows exponentially with the system size.
Another difficulty is that the spin-orbital interactions are generically
frustrated \textendash{} in the classical picture it's impossible
to find such a configuration that all the bonds have minimal energy,
and spins are entangled with orbitals, which means that the wave function
cannot be factorized into spin and orbital part \cite{Ole12}. This
means that an exact solution is not available (apart from very special
cases) and approximate solutions are limited by frustration and entanglement.
Nevertheless, in some parameters range, it's possible to find approximate
properties of the model (\ref{eq:sex}), such as spin and orbital
ordering or elementary excitations. There are Goodenough-Kanamori
rules saying that ferro/antiferromagnetic order in a given direction
is accompanied by antiferro/ferro order of orbitals (i.e., for $t_{2g}$
orbitals a pair of spins $\{\uparrow\downarrow\}$ on a given bond
is accompanied by a pair of orbitals $\{c,c\}$ and a pair of spins
$\{\uparrow\uparrow\}$ with a pair of orbitals $\{a,b\}$). However,
these are approximate rules that can be modified in presence of strong
quantum entanglement \cite{Ole06} or can be formulated in a more
general way \cite{Woh11}.

\begin{figure*}[t]
\begin{center}\includegraphics[width=0.8\textwidth]{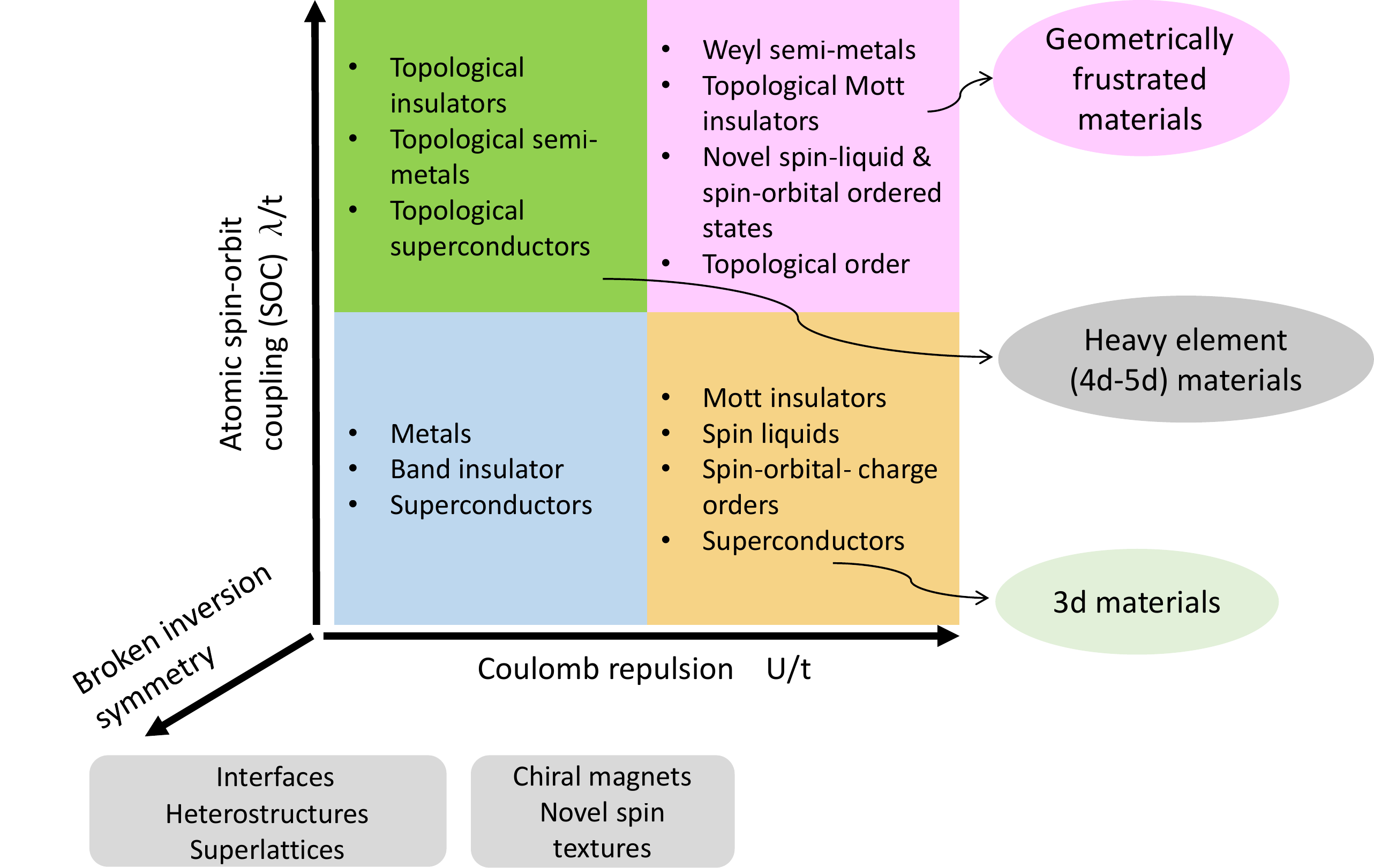}\end{center}\caption{Schematic view of possible states of matter of TMOs as function of
microscopic parameters; Coulomb repulsion $U$ and spin-orbit coupling
$\lambda$ with respect to hopping amplitude $t$. These quantities
can be changed by taking different transition metal ions, starting
from lighter and more correlated $3d$ metals up to more heavy $4d$
and $5d$ ones with less correlations and more SOC. An extra parameter
could be factors making the system non-centrosymmetric. \label{fig:2}}
\end{figure*}

Transition metal oxides are intensively studied because of broad variety
of possible non-trivial phenomena and possible states of matter that
can be realized in TMOs. This follows from the competition between
kinetic energy $\propto t$ of the electrons and various types of
ordering stabilized in the regime of large Coulomb repulsion $\propto U$.
Fig. \ref{fig:2} schematically depicts some of there states of matter
as functions of $U$ and spin-orbit coupling $\lambda$ with respect
to $t$. Low values of $\lambda$ are typical for $3d$ metals \cite{Ima98,Dag05},
for such compound as cuprates or iron pnicitides \cite{Nich11} known
for high-temperature superconductivity or manganites where colossal
magnetoresistance is observed. Depending on the interactions strength,
for small $U/t$ we can obtain weakly correlated states like metals
or band insulators or superconductor and for larger $U/t$ one gets
correlated insulators described by spin-orbital models (\ref{eq:sex}),
leading to various types of orders, like in $t_{2g}$ vanadates \cite{Kha01,Hor03,Ishi05},
$e_{g}$ manganites \cite{Yun98,Fei99,Dag04} and copper fluorides
\cite{Lak05,Fei97,Brz11-KKbi,Brz12,Brz13}, or to exotic spin-orbital
liquid or valence-bond phases on frustrated lattices \cite{Nor08,Rei05,Chal11}.
Apart from these orders strong correlations in $3d$ oxides can also
lead to supercoducting states \cite{Lee06,Wys17b} which can happen
as well in heavy fermion systems \cite{Wys16}. In the limit of larger
SOC, relevant for the $4d$ and $5d$ oxides \cite{Kim09,Pes10,Kre14,Kim14,Kha13},
one observes the so called topological states in the limit of small
interactions \cite{Has10} and topological order \cite{Lev06} in
the opposite limit \textendash{} the large $U/t$ typically requires
some frustration to reduce the hopping amplitude which is typically
larger than for the $3d$ oxides due to larger atomic shells. For
the $4d$ ruthenates this can be achieved by octahedral distortions
reducing the bandwidth and lead to different spin and orbital patterns
\cite{Cuo06,For10} that can be modified by a non-vanishing SOC \cite{Cuo06b}.
Another relevant degree of freedom in ruthenates is dimensionality,
controled by the layered structure of these compounds, that decides
about electronic properties of these systems \cite{Mal11,Mal13} and
can yield magnetic states with interlayer spin anisotropy \cite{Gra16}.
Another factor leading to exotic magnetic states can be breaking of
the inversion symmetry that can result in spin interaction terms of
the Dzyaloshinskii\textendash Moriya type $\vec{S}_{i}\times\vec{S_{j}}$
and chiral spin orders \cite{Bod06}. Such an effect can be achieved
by creating interfaces or heterostructures, as a monolayer-bilayer
ruthenate superlattice \cite{Aut14} or ruthenium-iridium oxide bilayer
hosting topological Hall effect \cite{Mat16,Brz18-ahe}. It is however
debated whether the source of this effect are so called magnetic skyrmions
triggered by the Dzyaloshinskii\textendash Moriya interactions \cite{Mat16}
or non-trivial topology of the underlying band structure related to
the spin-orbital fluctuations \cite{Brz18-ahe}. 

\begin{figure*}[t]
\begin{center}\includegraphics[width=0.8\textwidth]{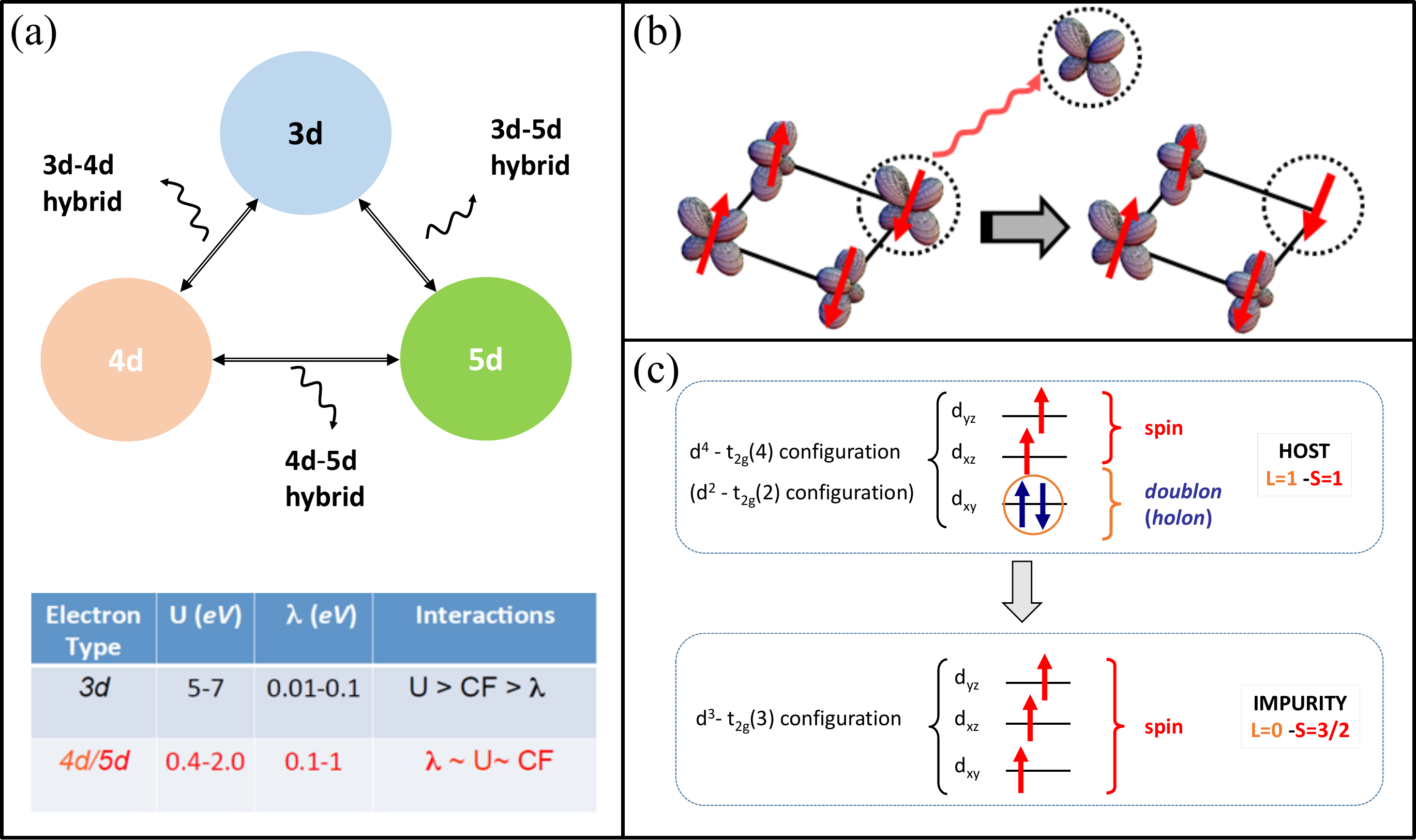}\end{center}\caption{(a) Schematic view of different oxide hybrids and typical energy scales
for the $3d$, $4d$ and $5d$ transition metals, i.e., Coulomb repulsion,
spin-orbit coupling and crystal field splitting. (b) Orbital dilution
by doping a $4d^{4}$ oxide with a $3d^{3}$ metal \textendash{} the
orbital degree of freedom effectively vanishes at doped lattice site.
(c) Atomic spin-orbital ground state for the atoms of the $4d^{4}$
(or $4d^{2}$) host and a $3d^{3}$ dopant. \label{fig:3}}
\end{figure*}

A new platform to obtain even more interesting phenomena and spin-orbital
orders could by hybrid oxides, i.e., such oxides where on random lattice
sites some metal ions are replace by other transition metals. In this
way one can create hybrids where energy scales of the Coulomb interaction,
spin-orbit coupling and crystal field splitting are not spatially
homogeneous, see Fig. \ref{fig:3}(a). An example of $3d-4d$ hybrids
are manganese doped layered ruthenium oxides from the family Sr$_{n+1}$Ru$_{n}$O$_{3n+1}$
($n$ defines number of layers), where physical properties strongly
depend on $n$. For $n\to\infty$, i.e., cubic compound, doping with
manganese drives the system from a metallic ferromagnet to antiferromagnetic
(AF) insulator \cite{Cao05}, whereas for $n=2$ the same doping gives
an insulator with a zigzag magnetic order \textendash{} spins order
ferromagnetically along zigzag lines in the planes \cite{Hos08,Mes12}.
For a similar $3d-4d$ hybrid, single-layer Ca$_{2}$RuO$_{4}$ oxide,
doping with chromium leads to noncollinear spin order with a tendency
to ferromagnetism and exotic negative volume thermal expansion in
the parameter range where the system exhibts spin and orbital order
\cite{Dur06,Qi10}. Doping of ruthenium oxides with $3d^{3}$ manganese
has a particularly simple interpretation in the limit of strong correlations
where the effective spin-orbital description (\ref{eq:sex}) is valid.
It is an effective dilution of the orbital degrees of freedom, as
shown in Fig. \ref{fig:3}(b)-(c), being a subject of papers \cite{Brz15X}
and \cite{Brz16}. Ruthenium atoms (being the host's atoms) are in
$4d^{4}$ configuration so their atomic state has spin $S=1$, according
to Hund's rule, and orbital angular momentum $L=1$, where orbital
degree of freedom is the double occupation (doublon) of one of the
$t_{2g}$ orbitals. On the other hand the doped ions (i.e., the impurities)
have $3d^{3}$ electronic configuration so their atomic state is the
one with maximal spin $S=3/2$. Since all the orbitals are occupied
by one electron each, effectively these ions have no orbital degrees
of freedom. In this context doping with chromium $3d^{2}$ is completely
different case because similarly to the host's atoms the dopants have
spin $S=1$ and orbital angular momentum $L=1$ realized by empty
occupation (holon) of one of the $t_{2g}$ orbitals \textendash{}
see Fig. \ref{fig:3}(c). Hence, there is no orbital dilution in this
case but the charge dilution \cite{Brz17}.

\begin{figure*}[t]
\begin{center}\includegraphics[width=1\textwidth]{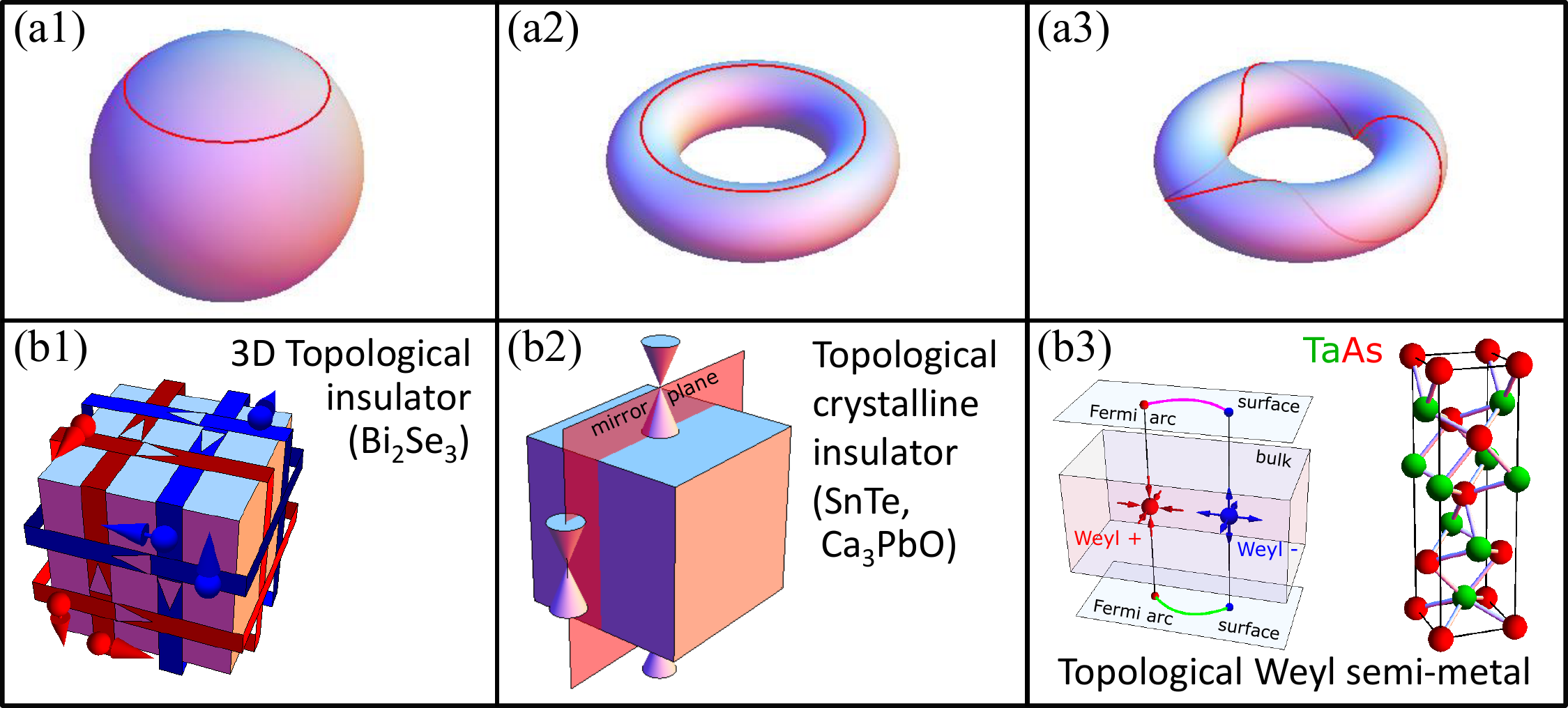}\end{center}\caption{Artist's view of the relation between topology and condensed matter
physics; (a1-a3) topological difference between torus and sphere described
by a loop on their surface, (b1-b3) examples of solids where non-trivial
topology of the bulk ground states leads to metallic edge states.
\label{fig:4}}
\end{figure*}

A completely different type of ordering, comparing to conventional
spin and orbital orders, is topological order. It refers to periodic
systems and can be observed when go around the system. For example,
running around the surface of a sphere, as shown in Fig. \ref{fig:4}(a1),
we do not observe anything particular because any loop on its surface
can squeezed to a point. On the other hand, doing the same on the
surface a torus, see Figs. \ref{fig:4}(a2-a3), we can notice that
loops can be non-trivially different; we can do $n$ laps along the
large circle and $m$ laps along the small one and we cannot go smoothly
between the loops with different $(m,n)$. This means that, in contrary
to a sphere, a torus has a non-trivial first homotopy group. This
follows of course from the fact that sphere and torus indeed have
different topology \cite{Hatch01}. This has consequences in the physics
of solids \textendash{} for a non-interacting fermion system the role
of loop is played by the quasimomentum space whereas the space of
eigenstates of a Hamiltonian plays the role of the surface on which
we trace the loop. Rigorously, such a link was established by the
Atiyah-Bott-Shapiro construction \cite{Ati64}, being rather advanced
mathematical concept. The main idea is that thanks to the global phase
invariance of the quantum states the eigenstates of Hamiltonians of
different symmetries can be regarded as isomorphic to the homogeneous
spaces of orthogonal, unitary and symplectic groups. Thus a given
Hamiltonian sets a maps from the Brillouin zone (BZ) to such a homogeneous
space, also called a classifying space. In contrary to multi-dimensional
spheres, for which complexity of higher homotopy groups is uncontrolable
\cite{Hatch01}, the homotopy structure of classifying spaces is strictly
limited by the so-called Bott periodicity \cite{Bott57},saying that
the homotopy groups are periodic in spatial dimension $D$ of the
BZ with period $2$ for unitary and $8$ for orthogonal and symplectic
cases. This allows to fully classify all possible topological states
of physical systems which is typically done by means of the so-called
algebraic K-theory, being a method of generalized cohomology groups
\cite{Hatch09}, a systematic but approximate approach that assumes
that the number of bands is infinite . Another important mathematical
property of the homogeneous spaces is that their topological properties
are independent of their dimension. This guarantees that the topology
of the bands does not depend on the choice of the unit cell, which
can be chosen as multiplicity of the elementary cell. 

Topological phases of matter can be characterized by topological invariants
that take integer values, in analogy to surfaces shown in Figs. \ref{fig:4}(a1-a3)
that can be characterized by integrals of the Gauss curvature giving
the number of holes in the surface ($0$ for sphere and $1$ for torus).
This is some kind of a quantization that has physical consequences
\textendash{} for example, in quantum Hall effect where it gives the
number of edge states and consequently quantization of the Hall conductance.
Depending on the basic (non-spatial) symmetries of the Hamiltonians,
which are time reversal symmetry ${\cal T}$, particle-hole ${\cal C}$
symmetry and chirality ${\cal S}$ (product of ${\cal T}$ and ${\cal C}$),
and spatial dimension of the system we can classify its topological
states as always trivial, non-trivial with a $\mathbb{Z}$ topological
index (any integer including $0$ being a trivial state) or non-trivial
with a $\mathbb{Z}_{2}$ index (only $0$ or $1$) \cite{Sch08,Shi14}.
Symmetries ${\cal T}$ (${\cal C}$) are antiunitary so they can be
either absent or present such that ${\cal T}^{2}=1$ (${\cal C}^{2}=1$)
or ${\cal T}^{2}=-1$ $({\cal C}^{2}=-1)$. On the other hand chirality
(or sublattice symmetry) is unitary so either the system has it or
not. All together, this gives ten canonical symmetry classes called
Altland-Zirnbauer classes, for which there exist six universal prescriptions
for calculating topological indices; chiral and non-chiral $\mathbb{Z}$
index and two types of $\mathbb{Z}_{2}$ index, both in chiral and
non-chiral versions. These apply to all Hamiltonians of non-interacting
fermions, both with energy gap (insulators, fully gapped superconductors)
\cite{Has10,Qi11} and without (semimetals, metals and nodal superconductors)
\cite{Zha13,Zha14}. In case of an energy gap we say that it is topologically
protected if the system has non-vanishing topological index \textendash{}
it is defined in fully energy-momentum space. On the other hand for
$D$-dimensional systems without energy gap we say that a Fermi surface
of dimension $d$ is topologically protected if it has a non-vanishing
topological charge \textendash{} it is a topological index defined
on a $p$-dimensional sphere enclosing Fermi surface in energy-momentum
space, where $p=D-d$.

From the existence of topological indices and charges one can derive
a very important property: the bulk-boundary correspondence. One can
show by a rigorous calculation that if a system have an edge (a boundary
with vacuum or other system) and it is topologically non-trivial (as
a periodic system without edge) then on this edge there will appear
states closing the energy gap or connecting the Fermi points in the
momentum space. In this way one can show that any change of the topological
index of a system must be related with closing the gap, if the system
has it, or opening it, if the system is gapless \cite{Ess11,Gur11}.
Since the topological index is given by an integer, a small perturbation
cannot change it, unless it changes the symmetries of a system, so
it cannot close or open the gap. This is meant by a topological protection
of a gap or a Fermi surface. These symmetries can be both non-spatial
(${\cal T}$ ,${\cal C}$ and ${\cal S}$) or spatial, related with
crystal symmetries like mirror symmetry, inversion with respect to
inversion center or rotation with respect to an axis, which additional
complicates classification of topological states \cite{Shi14,Chi14}.
Examples of topological systems are shown in Fig. \ref{fig:4}(b1-3).
These are; (b1) \textendash{} three-dimensional topological insulator
(TI) Bi$_{2}$Se$_{3}$ with metallic, so called helical edge states
\cite{Xia09}, (b2) \textendash{} topological crystalline insulator
(TCI), e.g., SnTe or Ca$_{3}$PbO protected by mirror reflection,
with Dirac fermions as the edge states \cite{Tan12,Woj14,Woj15} and
(b3) \textendash{} topological Weyl semimetal (SM) TaAs with Fermi
arcs on the surfaces \cite{Bur11,Xu15}. Following the bulk-boundary
correspondence the topological properties of the bulk are manifested
by the metallic egde states; in case of TI they experience spin-momentum
locling, i.e., the direction of motion determines direction of spin,
in case of TCI they form surface Dirac cones and in Weyl SM Fermi
arcs. Other example of topological semimetal is graphene \cite{Two06,Bar07,Net09}
hosting bulk Dirac fermions protected by mirror symmetry. Apart from
effective Dirac and Weyl particles low-dimensional topological systems
can also host so-called Majorana fermions, as in 1D SbIn superconducting
nanowires \cite{Lut10,Mou12,Sat16} or other SC systems with vortices
or defetcs \cite{Vol99,Wim10}, including even ultracold atom systems
\cite{Sat09}. Due to potential application in quantum computers \cite{Kit09},
for their non-Abelian braiding properties \cite{Hya13,Ful13}, Majorana
states are intensively searched in other than SC system, like quantum
Hall ferromagnets \cite{Kaz17,Sim18} and surfaces with atomic steps
\cite{Sessi16}. It is however not completely clear whether the zero-energy
modes observed at the surface steps of TCI \cite{Maz17} are related
to the Majorana quasiparticles or rather to the magnetic domain walls
crossing the steps \cite{Brz18-steps}.

General classifications of topological systems exist only for fermion
systems without interactions. Nevertheless, prescriptions for topological
invariants are applicable also in presence of interactions because
they are formulated in terms of single-particle Green's function.
Symmetries ${\cal T}$, ${\cal C}$ and ${\cal S}$ can be also generalized
for interacting systems. Hence, in principle one can show topological
non-triviality of a generic fermionic system but still this may not
lead to any edge states because interacting Green's function may have
not only poles but also zeros \cite{Ess11,Gur11}. In this sense interaction
can drive a topological system trivial, at least concerning the single
particle spectrum. In practice a good approximation of Green's function
in strongly correlated systems can be obtained by means of various
numerical implementations of Dynamical Mean Field Theory (DMFT) \cite{Bud13,Li15,Amma17}
or variational wave-function approaches \cite{Wys16b}. A slightly
different case is a Hamiltonian of the type (\ref{eq:sex}) with no
fermionic degrees of freedom, only interacting spins and orbitals.
In such case we can have a topological order which is a kind of a
non-local order protected by symmetries, i.e., symmetry-protected
topological phases \cite{Witt16}, with topologically robust ground-state
degeneracy . It may be not obvious whether a system have it. It is
known that in one dimension a system with topological order has degenerate
entanglement spectrum \cite{Pol12}. Another way to look for it is
by imposing twisted boundary conditions, parametrized by an angle,
and calculating Berry phase of the ground state acquired when changing
periodically this angle \cite{Hat05,Kar15}. A more involved approach
is by extracting so-called modular transformation matrices from the
degenerate ground-state wave functions obtained either by exact diagonalization
\cite{Claa17} or more involved tensor network type of approaches
\cite{Cin13}, including standard Density Matrix Reronalization Group
(DMRG) method \cite{Bal12}. These matrires encode self- and mutual-braiding
statistics of the elementary excitations \cite{Zha12,Zhu13} and are
closely related with entanglement spectra of a two dimensional systems
partitioned into halves along two different cuts. Such approach to
topological order can for instance help to identify topological spin-liquid
phases and fractional excitations of frustrated magnets \cite{Bal12,Claa17}.
\begin{figure*}[t]
\begin{center}\includegraphics[width=0.8\textwidth]{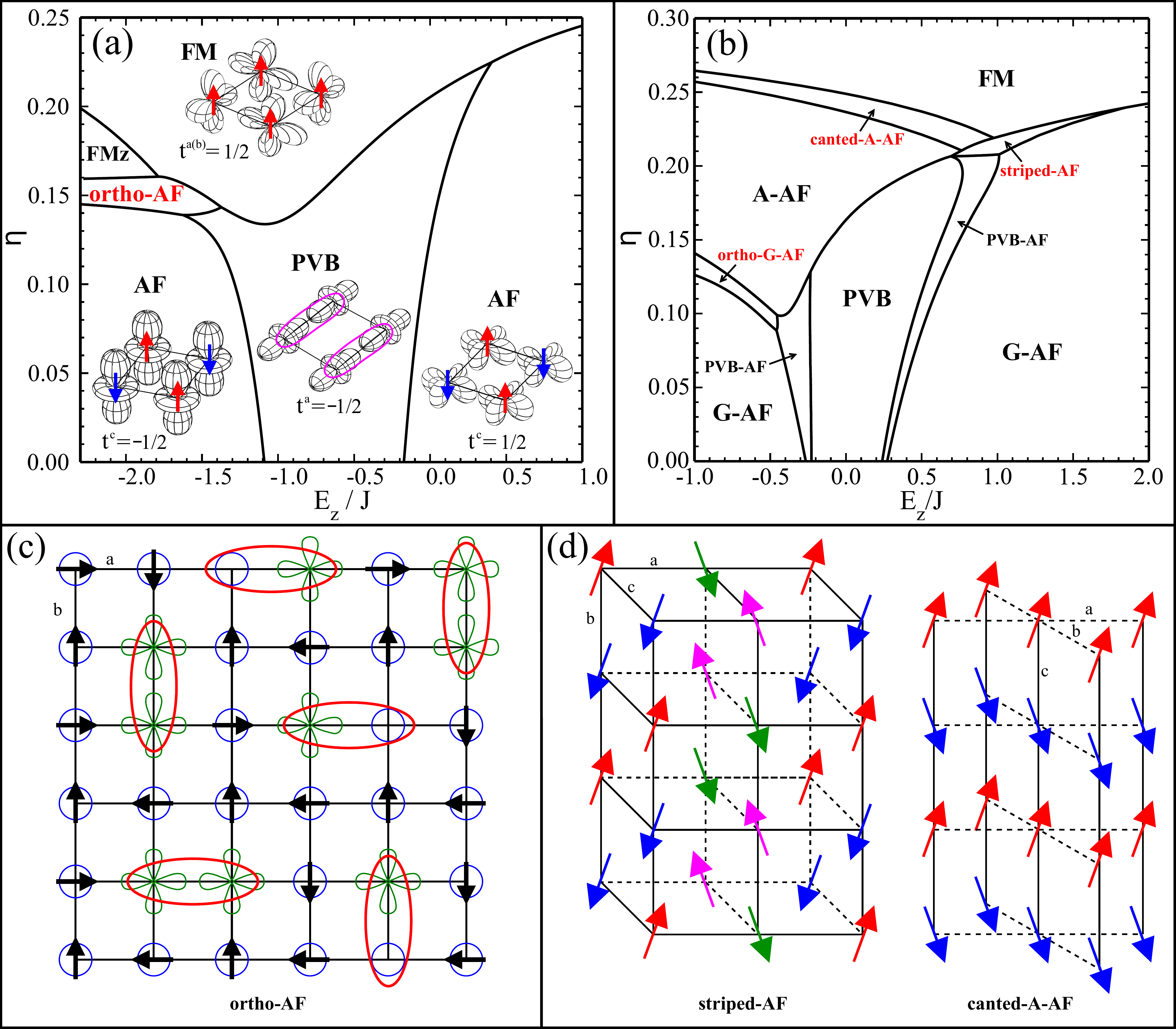}\end{center}\caption{Most relevant results for the spin-orbital Kugel-Khomskii model in
two and three dimensions. Phase diagrams as functions of crystal field
splitting $E_{z}/J$ and Hund's exchange $\eta=J_{H}/U$ for; (a)
two \cite{Brz12} and (b) three \cite{Brz13} dimensions. (c) Artist's
view of the ground state in ortho-AF phase in two dimensions, arrows
represent spins, circles and clovers are orbitals $|3z^{2}-r^{2}\rangle$
and $|x^{2}-y^{2}\rangle$, ellipses are spin singlets. (d) Schematic
view of the magnetic orders in striped-AF and canted-A-AF phase in
three dimensions with arrows being spins. \label{fig:5}}
\end{figure*}

In this paper we will addres the issue of spin, orbital and topological
order in the strongly correlated electron systems. In Sec. \ref{sec:Noncollinear-magnetic-order}
we focus on exotic cases of non-collinear magnetic order in a class
of the Mott insulating $d^{9}$ transition metal oxides which arises
in absence of the atomic SOC, in Sec. \ref{sec:Spin-orbital-model-with}
we show exotic case of an exact topological order in a one-dimensional
spin-orbital model that arises from orbitally degenerate Mott unsulators
and we link it with spontaneous dimerization due to orbital quatnum
fluctuations., in Sec. \ref{sec:Inhomogeneous-spin-orbital-model}
we discuss the impact on spin and orbital order of orbital and charge
dilution in inhomogeneous (hybrid) $d^{3}-d^{4}$ and $d^{2}-d^{4}$
transition metal oxides and the quantum aspect of orbital dilution
are discussed in Sec. \ref{sec:Orbital-dilution-in}. Sec. \ref{sec:Magnetic-zigzag-phases}
addresses the question of stability of zigzag antiferromagnetic phases
in a bilayer and monolayer $d^{3}-d^{4}$ hybrid oxide and in Sec.
\ref{sec:Topological-semi-metal-phases} the topological properties
of such phases are discussed with special focus on the role of the
nonsymmorphic symmetries. Finally, Sec. \ref{sec:Topological-phases-in}
describes topological properties of a non-uniform Kitaev model in
one dimension which originates from spin-orbital model of a $d^{2}-d^{4}$
hybrid oxide, see Sec.\ref{sec:Relationship-of-non-uniform}. The
summary is given in Sec. \ref{sec:Summary}.

\section{Noncollinear magnetic order stabilized by orbital fluctuations\label{sec:Noncollinear-magnetic-order}}

A very interesting feature of the Kugel-Khomskii model (\ref{eq:sex})
for strongly correlated TMOs with the $d^{9}$ electron configuration
of the metal ions are exotic spin orders found in two and three dimensions
\cite{Brz12,Brz13}. The $d^{9}$ configuration effectively means
a single hole in the multiplet of $d$ orbitals, so spin $S=1/2$
and orbital degrees of freedom realized by a hole occupying one of
$e_{g}$ orbitals yielding orbital pseudospin $T=1/2$. Considered
models are for two-dimensional (2D) \cite{Brz12} and three-dimensional
(3D) \cite{Brz13} cubic lattices relevant for description of copper
compounds, namely K$_{2}$CuF$_{4}$ and KCuF$_{3}$ (with analogical
structure as TMOs). Study of these compounds is qualitatively simplified,
and can be limited to spin-orbital superexchange, because SOC is absent.
Nevertheless it revealed peculiar types of spin ordering, namely noncollinear
magnetic patterns depicted in Figs. \ref{fig:5}(c-d). The mechanism
that can stabilize such ordering in absence of SOC are entangled spin-orbital
fluctuations on lattice bonds, also dicussed in Refs. \cite{Brz12-KK2,Brz12-KK,Brz14-KK}. 

Noncollinear phases were observed in spin-orbital phase diagrams,
obtained via cluster mean-field method (Bethe-Peierls-Weiss method,
also used in analogical bilayer case \cite{Brz11-KKbi}), shown in
Figs. \ref{fig:5}(a-b). The parameters are; crystal field splitting
$E_{z}$ with respect to superexchange constant $J=4t^{2}/U$ (where
$t$ is a hopping amplitude for a pair of orbitals $|3z^{2}-r^{2}\rangle$
along $c$ axis), where positive value of $E_{z}$ means that orbital
states $|x^{2}-y^{2}\rangle$ have lower energy than $|3z^{2}-r^{2}\rangle$
and negative vice versa, and value of $\eta=J_{H}/U$ being the ratio
of Hund's exchange with respect to Hubbard $U$. Main magnetic phases
are those with AF order in all directions (labeled as AF in 2D and
G-AF in 3D systems), phases with ferromagnetic (FM) order in all direction
(labeled as FM) and in 3D case a phase being FM in $ab$ planes and
AF in $c$ direction (labeled as A-AF). In all these phases the Goodenough-Kanamori
rules are satisfied meaning that orbital order in AF bond directions
is ferro-orbital and vice versa \textendash{} see Fig. \ref{fig:5}(a).
An exception is the FMz phase in 2D system where orbitals do not alternate
despite FM spin order. One can also notice that the increasing value
of $\eta$ decides about the tendency towards ferromagnetism whereas
$E_{z}$ determines orbital polarization. These relatively simple
phases, whose existence can be predicted in the classical limit, are
not the only ones. In case when energy scales compete and the system
cannot decide about any simple type of order, phases with strong quantum
fluctuations occur, such as plaquette valence bond (PVB) phase or
finally phases with noncollinear spin order, labeled in red in diagrams
\ref{fig:5}(a-b). These are phases; ortho-AF in a 2D system and ortho-G-AF,
canted-A-AF and striped-AF in three dimensions schematically depicted
in Figs. (c-d).

It was demonstrated that the ground state in the ortho-AF phase in
the classical limit, see Fig. \ref{fig:5}(c), consists of mutually
perpendicular NN spins and all orbitals in $|3z^{2}-r^{2}\rangle$
states \cite{Brz12}. This state is dressed with quantum fluctuations
having a form of singlets accompanied with a pair of orbitals $|x^{2}-y^{2}\rangle$
or with a single $|x^{2}-y^{2}\rangle$ orbital. Noncollinear spin
configuration in this case is related with entanglement of spin and
orbital degrees of freedom on the bonds, which can be proven by a
perturbative expansion. Taking the crystal field term $\propto E_{z}$,
a purely orbital Hamiltonian, as a center of the expansion and the
rest of the Hamiltonian (including spins) as a perturbation, one can
get in the first order effective spin interactions between second
neighbors on the square lattice. Thus the two sublattice are decoupled
and order antiferromagnetically, one independently of the other. The
orbital states remains unchanged. In the second order the orbital
fluctuations occur, as shown in Fig. \ref{fig:5}(c), and effective
four-spin interactions which couple the two sublattices in the way
that, after neglecting quantum fluctuations, the neighboring spins
orient themselves perpendicularly. 

This mechanism, described in work \cite{Brz12}, is a new way to obtain
a noncollinear magnetic order without involving strong relativistic
effects, present in heavy transition metals. The key ingredient are
strong orbital fluctuations, which in this case follow from the proximity
of an orbital phase transition from ferro-orbital order in AF phase
to alternating orbitals in the FM one. A similar effect can be also
observed in a 3D system, where the ortho-G-AF phase is realized, being
a 3D analogue of ortho-AF \cite{Brz13}, together with canted-A-AF
and striped-AF phases, depicted in Fig. \ref{fig:5}(d). They all
appear as noncollinear intermediate phases between phases with conventional
spin order. For example, being in the phase diagram \ref{fig:5}(b)
in phase G-AF on the left, spin correlations are AF in all directions
and orbitals are polarized as $|3z^{2}-r^{2}\rangle$. Increasing
$\eta$ we change spin correlations in the $ab$ planes to FM ones
and orbitals to partially alternating by passing to the ortho-G-AF
phase. By further increasing $\eta$ we change all the spin correlations
in FM ones, so the bonds in $c$ direction must reorient themselves
by $180$ degrees. This reorientation takes place in the canted-A-AF
phase, see Fig. \ref{fig:5}(d), where the angle between spins along
$c$ axis changes continuously from $180$ to $0$ degrees by increasing
$\eta$. This process can be described by a similar perturbative expansion
as in the ortho-AF phase with a difference that the center of the
expansion in the part of the Hamiltonian that favors alternating orbitals
\cite{Brz13}. On the other hand, the striped-AF phase, also shown
in Fig. \ref{fig:5}(d), can be obtained by the same expansion as
ortho-AF one but with positive crystal field $E_{z}$, meaning a state
with polarized $|x^{2}-y^{2}\rangle$ orbitals. The result is then
completely different, the spin correlations are AF in all directions
but there is a deviation from the $180$ degree angle between neighboring
spins along one lattice direction and this gives a four sublattice
magnetic order. Therefore, it was demonstrated that complex spin and
orbital orders can arise even in TMOs with negligible SOC.

\begin{figure*}[t]
\begin{center}\includegraphics[width=0.8\textwidth]{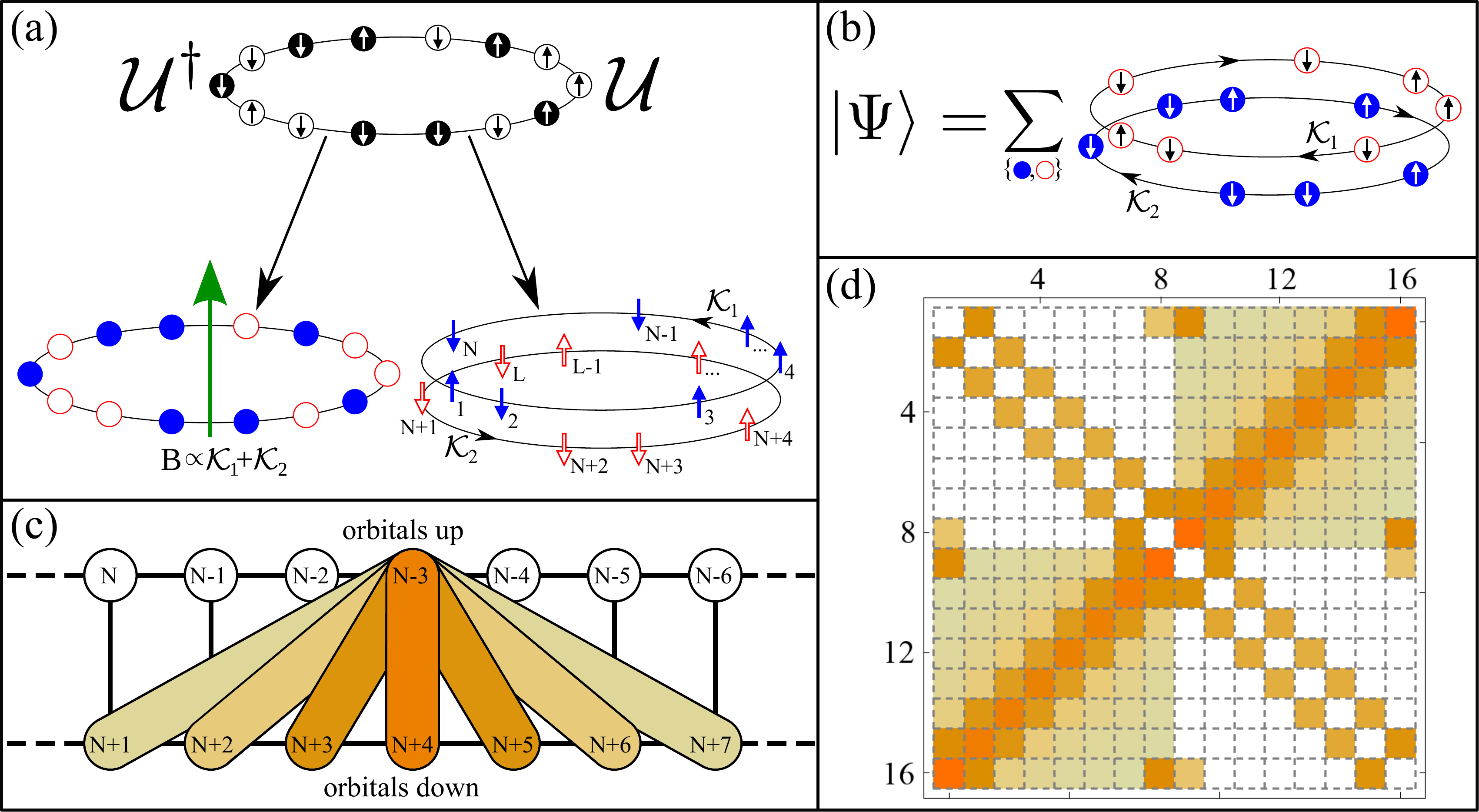}\end{center}\caption{Qualitative results for spin-orbital SU$(2)\otimes XY$ model without
(a-b) and with Heisenberg term (c-d) for spin-spin interactions. (a)
Schematic view of spin-orbital splitting realized by ${\cal U}$ transformation.
Initial spin-orbital ring (top) in split into purely orbital (left)
and purely spin (right) part. (b) Schematic view of the ground state.
States with a fixed number o orbitals 'up' (empty circles) and \textquoteright down\textquoteright{}
(full circles) form an orbital Fermi sea (sum over orbitals). Every
such state is dressed with spin currents ${\cal K}_{1,2}$, flowing
through subsystem of orbitals 'up' and 'down'. (c) Schematic view
of effective spin bonds between two subsystems (two legs of the ladder)
in the SU$(2)\otimes XY$ model with Heisenberg term treated perturbatively,
where $N=L/2$ and $L$ is the system size. The marked bonds connect
site $l=4$ of subsystem with orbitals 'up' with sites of the subsystem
with orbitals 'down'. Their color saturation reflects strength of
the effective $J_{ij}$ (always $J_{ij}\geq0$). (d) Matrix of the
effective spin bonds $J_{ij}$ for $L=16$, the color scale as in
panel (c).\label{fig:6}}
\end{figure*}

\section{Spin-orbital model with topological order and spontaneous dimerization\label{sec:Spin-orbital-model-with}}

Even more exotic order is present in one-dimensional (1D) spin-orbital
${\rm SU}(2)\otimes{\rm XY}$ model introduced by Kumar \cite{Kum13}
in the context of spin-orbital separation found in 1D TMO \cite{Schla12}.
Such models are challenging in low dimension as they often exhibit
strong spin-orbital entanglement \cite{You15b} and quantum critical
points \cite{You15}. The model has a generic form of equation (\ref{eq:sex})
with spins $S=1/2$ and orbital pseudospins $T=1/2$, so we will use
notions spins up and down but also orbitals 'up' and 'down'. To some
extent it resembles exatly solvable orbital compass models \cite{Dou05,Tro12,Brz10-compa,Brz10-hidim,Brz13-compa2d,Nuss15}
in one dimension \cite{Brz07-compa,Brz09-acta} and on a ladder \cite{Brz09-colad,Brz08-epjb}
that originate from interacting orbital models. Their generalization
are so-called compass plaquette models \cite{Wen09}, that in one
dimension can be also solved exactly, at least in certain limits \cite{Brz14-plaq,Brz16-plaq2},
making use of specific local symmetries. These models however contained
only orbital pseudospins and no real spins. The case of the ${\rm SU}(2)\otimes{\rm XY}$
model is different; Kumar has demonstrated thar for an open chain
one can define a unitary transformation ${\cal U}$ that splits spin
and orbital degrees of freedom in the rigorous way. The effect of
this transformation is that spins become an effective gauge field
attached to orbitals and after the transformation they disappear completely
from the Hamiltonian. However, in Ref. \cite{Kum13} there is no answer
to the question what happens in the system has periodic boundary conditions
(PBCs). This turns out to be particularly interesting \textendash{}
in work \cite{Brz14} it was shown that transformation ${\cal U}$
still leads to almost complete spin-orbital separation but the spins
absorbed as a gauge field by orbitals reappear on the last bonds connecting
first and the last site of the chain. They appear as a very special
non-local operator leading to topological order and topological excitations
in the system. This is an intriguing example of a topological order
in a strongly correlated system that is exactly solvable. 

The action of the non-local boundary term on the spin subsystem is
generating cyclic transpositions, i.e., every spin at site with orbital
'up' ('down') is shifted right to the nearest site with orbital 'up'
('down'), and thereby the total number of orbitals 'up' ('down') is
a good quantum number. On the other hand the orbital subsystem feels
these cyclic transpositions as effective magnetic field crossing the
closed chain \textendash{} ring. A schematic view of such a special
spin-orbital splitting caused by transformation ${\cal U}$ is shown
in Fig. \ref{fig:6}(a) and of the ground state $|\Psi\rangle$ with
topological order in Fig. \ref{fig:6}(b). The position of orbitals
'up' and 'down' in such a state fluctuates in the orbital Fermi sea
in such a way that on average every 'up' orbital is neighboring with
'down' one. At the same time in every component of this superposition
there are spin currents with quasimomenta ${\cal K}_{1}$ i ${\cal K}_{2}$
flowing through subsystems of orbitals 'up' and 'down'. Thus, the
spin order in state $|\Psi\rangle$ is completely non-local and relies
on closed topology of the periodic chain \textendash{} hence it is
a topological order \cite{Brz14}. 

The minimal energy of state $|\Psi\rangle$ is obtained for ${\cal K}={\cal K}_{1}+{\cal K}_{2}=0$
and the lowest excitations are for ${\cal K}\not=0$, with excitation
energy being quadratic in ${\cal K}$ and energy gap between excited
states scales as $L^{-3}$, where $L$ is the system size. Such excitation
can be called topological, unlike orbital excitation whose gap scales
as $L^{-1}$. In the case of an open chain the states split by finite
${\cal K}$ collapse on each other to a single multiplet with degeneracy
$2^{L}$. One can see then that in the ${\rm SU}(2)\otimes{\rm XY}$
model topology determines the degeneracy $d$ of the ground state,
which for large system is $d\simeq2^{L+1}/L$ for closed chain and
$d=2^{L}$ for open one. In case of topological insulators such a
change of degeneracy follows from the presence of edge states with
zero energy. However, the ${\rm SU}(2)\otimes{\rm XY}$ model is not
a free fermion model and does not have single-particle states so it
is hard to talk about edge states in this context. Nevertheless the
topologically protected degeneracy and non-local spin order define
the topological order in this case.

The ${\rm SU}(2)\otimes{\rm XY}$ model has a rather special form
without any direct interaction between spins, only interaction between
spin-orbital pairs on the neighboring sites are present. It is then
natural to ask what will happen with the ground state if we add as
a small perturbation Heisenberg interactions between spins and what
kind of order it will produce \cite{Brz15}. The key question is to
express the pure spin Hamiltonian of the perturbation in the basis
defined by transformation ${\cal U}$ that causes spin-orbital splitting,
as shown in Fig. \ref{fig:6}(a), and deriving its effective form
by perturbation expansion in the spirit of works \cite{Brz12,Brz13}.
As an effect of this approach we obtain a ground state with dimerized
spin order \textendash{} NN correlation $\langle\vec{S}_{i}\vec{S}_{i+1}\rangle$
alternate between low and high values \cite{Brz15}. This is a non-trivial
result because in the initial Hamiltonian all the bonds have the same
strength. The exotic spin order follows here from the orbital fluctuations,
similarly as it was in the Kugel-Khomskii model \cite{Brz12,Brz13}.
These fluctuation lead to an effect known in the context of electron-phonon
interactions as Peierls dimerization \cite{Brz15} but here it happens
at zero temperature, in contranst to spin-orbital models studied before
where it was activated thermally \cite{Sir08}.

The mechanism of dimerization in the ${\rm SU}(2)\otimes{\rm XY}$-Heisenberg
model is related with the special action of transformation ${\cal U}$
on spin degrees of freedom. Under its action spins become so to say
attached to orbitals, so delocalization of orbitals entails delocalization
of spins. Spins are no longer associated to the lattice sites but
to the position of the first, second, third etc. 'down' orbital. In
this way we obtain an effective spin Hamiltonian of the form ${\cal H}_{{\rm eff}}=\sum_{i,j}J_{ij}\vec{S}_{i}\vec{S}_{j}$,
where due to delocalization of orbitals couplings $J_{ij}$ do not
refer only to NNs but they are diffuse as shown in Figs. \ref{fig:6}(c)
and \ref{fig:6}(d). Then, we have in Fig. \ref{fig:6}(c) a subsystem
of spins $\{\vec{S}_{1},\vec{S}_{2},\dots,\vec{S}_{N}\}$ attached
to orbitals 'up' (where $N=L/2$) and a subsystem of spins $\{\vec{S}_{N+1},\vec{S}_{N+2},\dots,\vec{S}_{2N}\}$
attached to orbitals 'down' depicted as two legs of a ladder. Dominating
couplings $J_{ij}$ are those on the rungs of the ladder \textendash{}
if the orbital state was a classical N\'eel state then it would be
the only non-vanishing $J_{ij}$. Due to orbital fluctuations we also
have couplings to the further neighbors on different leg of the ladder
and couplings along the legs for NNs, not shown in Fig. \ref{fig:6}(c)
\textendash{} all values of $J_{ij}$ for a system of the size $L=16$
are shown in Fig. \ref{fig:6}(d) (presence of finite $J_{1,N}$ follows
from PBCs). Dominating couplings are on antidiagonal of matrix $\mathbf{J}$
and for the ground state it is not a bad approximation to take only
these terms in the effective Hamiltonian ${\cal H}_{{\rm eff}}$.
The spin order that we get then in the physical basis, after inverse
${\cal U}$ transformation, is a dimerized state where on every odd
bond we have $\langle\vec{S}_{i}\vec{S}_{i+1}\rangle<0$ and on every
even one $\langle\vec{S}_{i}\vec{S}_{i+1}\rangle=0$. 

These perturbative considerations have limited application in the
thermodynamic limit $L\to\infty$ because the energy gap in the (open)
system vanishes as $L^{-1}$. Thus, for large system a numerical approach
was required, going beyond the limitations of perturbation theory.
The most suitable method for a 1D system is the density matrix renormalization
group (DMRG) \cite{Whi92}, which allows in this case to solve the
systems up to $L=600$ sites. These calculations showed that, quite
contrary to perturbative intuition, there is a continuous quantum
phase transition between a system with zero and positive (AF) Heisenberg
coupling between the spins. What is interesting is that a closed system
realizes a resonating state, being a superposition of two equivalent
dimerized states, which does not break a translational invariance
and which is subject to spontaneous symmetry breaking in the limit
of $L\to\infty$. On the other hand a perturbative effect, but of
higher order than ${\cal H}_{{\rm eff}}$, is dimerization of the
orbital state, being a consequence of spin dimerization, observed
in the DMRG calculations. All these exotic quantum effects are observed
only in case of AF Heisenberg exchange, whereas for FM coupling the
spin ground state is trivial and purely classical. Thus, for the dimerization
effect one needs entanglement of spin and orbital degrees of freedom.

\section{Inhomogeneous spin-orbital models; orbital and charge dilution\label{sec:Inhomogeneous-spin-orbital-model}}

\begin{figure*}[t]
\begin{center}\includegraphics[width=0.8\textwidth]{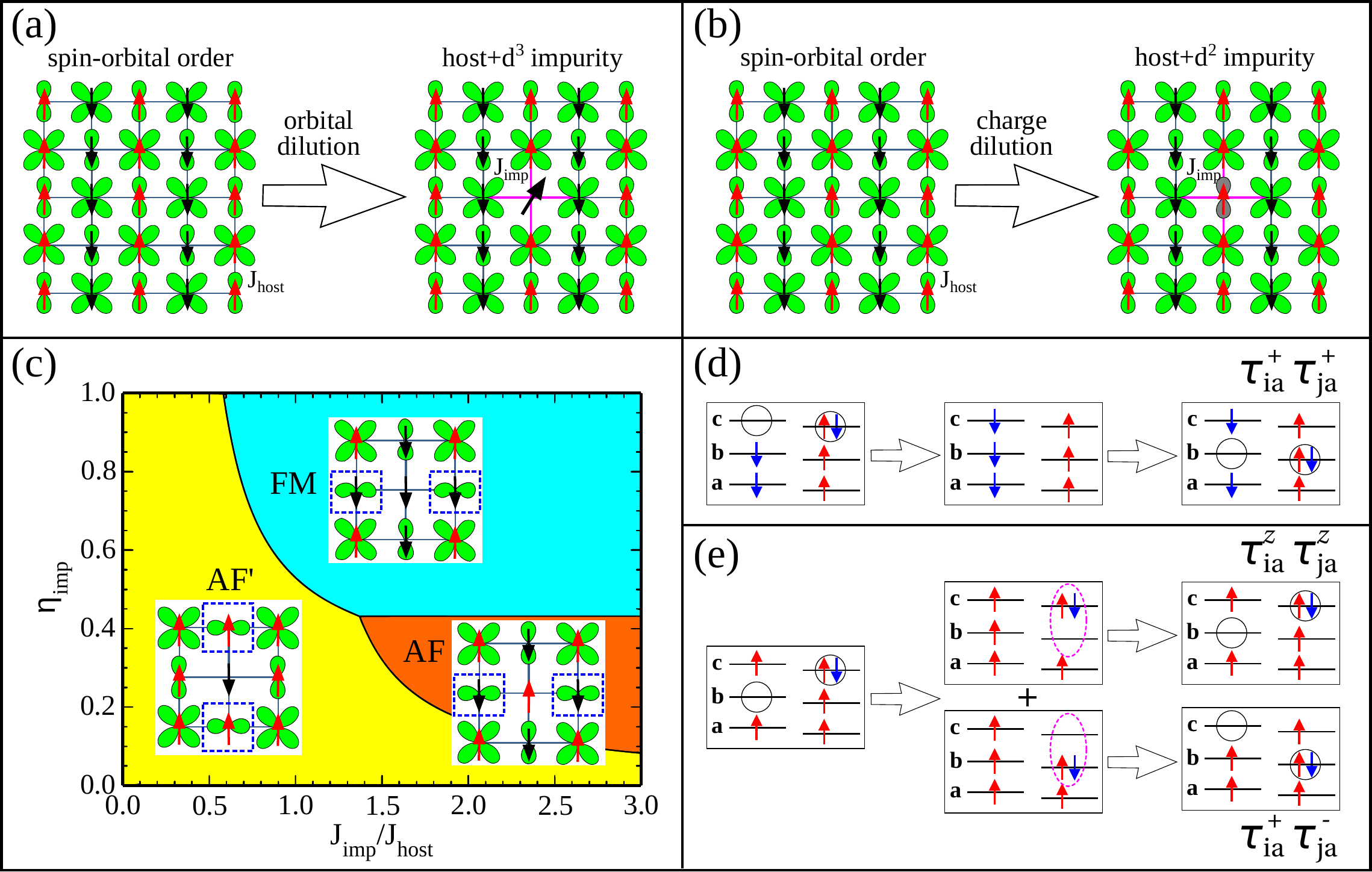}\end{center}\caption{Top \textendash{} doping of the transition metal ion in the $4d^{4}$
system with C-AF spin-orbital order with alternating $a$ i $c$ orbitals
(green shapes). (a) Orbital dilution due to $d^{3}$ doping (only
spin $S=3/2$, big arrow) i (b) charge dilution due to $d^{2}$ doping
(the same spin $S=1$ as host and and angular momentum $L=1$ realized
by a holon in orbital shaded gray). Bonds $J_{{\rm imp}}$ of impurity
with host (hybrid bonds) are marked in red and bonds $J_{{\rm host}}$
inside the host in black. Bottom left, (c) classical phase diagram
for a single $d^{3}$ impurity at host site where doublon was in orbital
$c$; frame marks the change of orbital or spin state around the impurity.
Bottom right, orbital fluctuations on hybrid bond $d^{2}-d^{4}$ with;
(d) AF and (e) FM spin correlations. Ellipse in the intermediate state
means delocalization of the doublon due to Hund's exchange. Reproduced
with permission from \cite{Brz17}. Copyright 2017 IOPscience.\label{fig:7}}
\end{figure*}

Up to now the models that were considered concerned only homogeneous
systems and without SOC. Despite this apparent simplicity it was still
possible to obtain interesting spin and orbital orders including topological
order in one dimension. Thus, even more interesting and richer perspective
would be due to hybrid systems like layered ruthenium $4d^{4}$ oxides
doped with manganese $3d^{3}$. Such a doping, as pointed out in the
introduction, is a dilution of orbital degrees of freedom, depicted
in Figs. \ref{fig:3}(b) and \ref{fig:7}(a). 

In the limit of strong correlations an effective spin-orbital model
of the type (\ref{eq:sex}) for bonds connecting the atoms of the
$d^{4}$ host is known. It contains interacting spins $S=1$ and orbital
pseudospins $T=1$, describing a doublon in one of the three $t_{2g}$
orbitals, as in Fig. \ref{fig:3}(c). On the other hand, for hybrid
bonds between $d^{4}$ host sites and $d^{3}$ impurities, the superexchange
Hamiltonian describing interactions between a pair spin-pseudospin
$S=T=1$ and a single spin $S=3/2$ was not known before. Its derivation
is given in Ref. \cite{Brz15X} as one of the main results. The main
issued addressed there is the change of spin and orbital order due
to doping of the host with quite typical spin-orbital C-AF order,
shown in Fig. \ref{fig:7}(a). These studies concerned both a single
impurity and finite doping case with periodic distribution of impurities.
For both these cases phase diagrams, as the one shown in Fig. \ref{fig:7}(c),
were determined containing different spin and orbital orders around
the dopants as function of microscopic parameters of the model. An
extension of this work for different configurations and concentrations
of impurities was presented in Ref. \cite{Brz16}. The interesting
point about both these works is that they show how by a purely magnetic
dopant one can affect both spin and orbital order.

Hybrid bonds around the impurities are significantly different than
the ones of the host. In case of the host bonds virtual processes
that lead to spin-orbital interactions engage charge excitations related
to interaction $U$, i.e., the lowest excited $d^{3}-d^{5}$ states
of a pair of ions $d^{4}-d^{4}$ have energy gap of the order of $U$.
In case of hybrid bonds of pair of ions $3d^{3}-4d^{4}$ the energy
gap to the lowest excited states is not given by $U$ but by ionization
energy $I_{e}$, being a bare energy difference between the $d$ levels
of manganese and ruthenium ions, following for instance from differing
main quantum numbers. As shown in Ref. \cite{Brz15X} this bare difference
in the excited $3d^{4}-4d^{3}$ states is additionally dressed by
differences in Hubbard and Hund's interactions between $3d^{3}$ and
$4d^{4}$ ions. Thus, we obtain the energy scale of the charge excitations
as $\Delta=I_{e}+3(U_{1}-U_{2})-4(J_{1}^{H}-J_{2}^{H})$, where $U_{1(2)}$
and $J_{1(2)}^{H}$ are Hubbard and Hund's interactions for $3d^{3}$($4d^{4}$)
ions. The superexchange constant $J$ for host bonds is then given
by a standard $J_{{\rm host}}\propto t_{{\rm host}}^{2}/U_{2}$, where
$t_{{\rm host}}$ is the hopping amplitude between $4d^{4}$ ions
(in fact this is a product of two hopping amplitudes, from ruthenium
to oxygen and from oxygen to ruthenium). On the other hand, for hybrid
bonds the superexchange constant is given by $J_{{\rm imp}}\propto t_{{\rm imp}}^{2}/\Delta$,
where $t_{{\rm imp}}$ is the hopping amplitude between $4d^{4}$
and $3d^{3}$ ions. For host bonds the parameter that decides about
tendency to ferro/antiferromagnetism is, as in the case of Kugel-Khomskii
model for $d^{9}$ ions \cite{Brz12,Brz13}, the ratio $\eta_{{\rm host}}=J_{2}^{H}/U_{2}$.
For hybrid bonds analogical role is played by $\eta_{{\rm imp}}=J_{1}^{H}/\Delta$. 

Figure \ref{fig:7}(c) shows a phase diagram, obtained by neglecting
quantum fluctuations, as a function of $J_{{\rm imp}}/J_{{\rm host}}$
and $\eta_{{\rm imp}}$ (with fixed $\eta_{{\rm host}}=0.1$). Single
impurity placed at the lattice site where, before doping, there was
a doublon of the host in orbital $c$ couples with its spin either
ferro or antiferromagnetically with surrounding spins of the host
(phases FM, AF or AF') and at the same time either polarizes the host's
orbitals 'towards itself' (phases FM and AF) or it is ignored by them
(phase AF'). Such defects resemble spin-orbital polarons considered
in vanadates where a doped hole is strongly localized at the charge
Ca defect and forms a spin-orbital polaron around it \cite{Ave15,Ave18,Ave18b}.
This is a different case than a hole doped in spin-orbital systems
\cite{Dag08,Woh09,Ber09,Bie17} that can delocalize due to spin quantum
fluctuations. Polarization of host's orbitals is possible when $J_{{\rm imp}}$
is sufficiently large with respect to $J_{{\rm host}}$, which also
depends on $\eta_{{\rm imp}}$. It follows from the fact that host's
orbitals pointing towards the impurity increase kinetic exchange between
$4d^{4}$ and $3d^{3}$ ions. In Ref. \cite{Brz15X} it was shown
that in case of finite doping there is a generic intermediate phase
between FM and AF phases where the impurity spin is frustrated, i.e.,
half of the hybrid bonds if FM and half is AF. Such a phase is denoted
as FS \textendash{} frustrated spin.

Other type of doping of the $4d^{4}$ system with C-AF order is doping
with $4d^{2}$ ions, as shown in Fig. \ref{fig:7}(b). In this case
both host and impurity are described by local spins $S=1$ and orbital
pseudospins $T=1$ but the charge related to the orbital degrees of
freedom is different. In case of host's ions we have doublons, i.e.,
double occupations of orbitals $a$, $b$ or $c$, and in case of
impurity we have holons being empty occupations. Thus we call such
a doping a charge dilution, in contrast to orbital dilution described
earlier. 
Hybrid bonds between host and impurity in the superexchange limit
require deriving, like in the case of $3d^{3}$ doping, and it is
done in Ref. \cite{Brz17}. The most interesting effect of charge
dilution is appearance of the orbital pairing terms of the form $\tau_{i\gamma}^{+}\tau_{j\gamma}^{+}$
in the spin-orbital Hamiltonian around the impurity. These terms are
absent in the pure host system and do not appear in case of orbital
dilution. Operators $\vec{\tau}_{i\gamma}$ are represented by the
three Pauli matrices on site $i$, which for a bond in the direction
$\gamma=a$, act in the space of orbitals states $|b\rangle$ and
$|c\rangle$ as if they were states of spin $S=1/2$, and analogically
in other directions $\gamma$. A consequence of presence of orbital
pairing terms is that 'orbital magnetization', being a total number
of doublons/holons in orbitals $a$, $b$ and $c$ in the system,
is not a good quantum number, similarly as in a superconducting Hamiltonian
total number of electrons is not a good quantum number. Quantum fluctuations
in the orbital sector are thus locally enhanced by the doping which
depending on the initial order of the host may affect or not the global
orbital order \cite{Brz17}. Other consequence of orbital pairing
$\tau_{i\gamma}^{+}\tau_{j\gamma}^{+}$ is that in 1D case for the
orbital sector we get a model equivalent to the $p$-type superconducting
Hamiltonian or Kitaev model \cite{Kit09}. Such a model is known for
its topologically non-trivial ground state which suggests that the
orbital state of a 1D $d^{4}-d^{2}$ model can also be topological
\textendash{} this problem is considered in Ref. \cite{Brz17doi},
which will be discussed later. This is a quite unexpected and interesting
aspect of charge dilution.

Exemplary virtual processes leading to orbital pairing terms are shown
in Fig. \ref{fig:7}(d); we assume bond in $a$ direction, so the
only allowed hopping processes are from orbital $b$ to $b$ and from
$c$ to $c$. In initial configuration the total spins on $d^{2}$
and $d^{4}$ ions are opposite and both doublon and holon are in orbitals
$c$. This an AF and ferro-orbital configuration. In virtual excited
state one of the electrons forming the doublon recombine with holon
so there is neither doublon nor holon in this state. Coming back to
the ground state one can now either reverse the previous process or
shift an electron from orbital $b$ of the impurity back to the orbital
$b$ of the host. In the latter case we come back to the ground state
where both holon and doublon were simultaneously shifted from orbitals
$c$ to $a$. In the language of orbital pseudospins $\vec{\tau}_{ia}$
this means that an operator $\tau_{ia}^{+}\tau_{ja}^{+}$ acted on
initial ground state. Of course, for a hybrid bond there are other
processes possible, shown in Fig. \ref{fig:7}(e), that lead to 'orbital
hopping' terms $\tau_{ia}^{+}\tau_{ja}^{-}$ or diagonal $\tau_{ia}^{z}\tau_{ja}^{z}$
ones, both also present on the host bonds.

\begin{figure*}[t]
\begin{center}\includegraphics[width=1\textwidth]{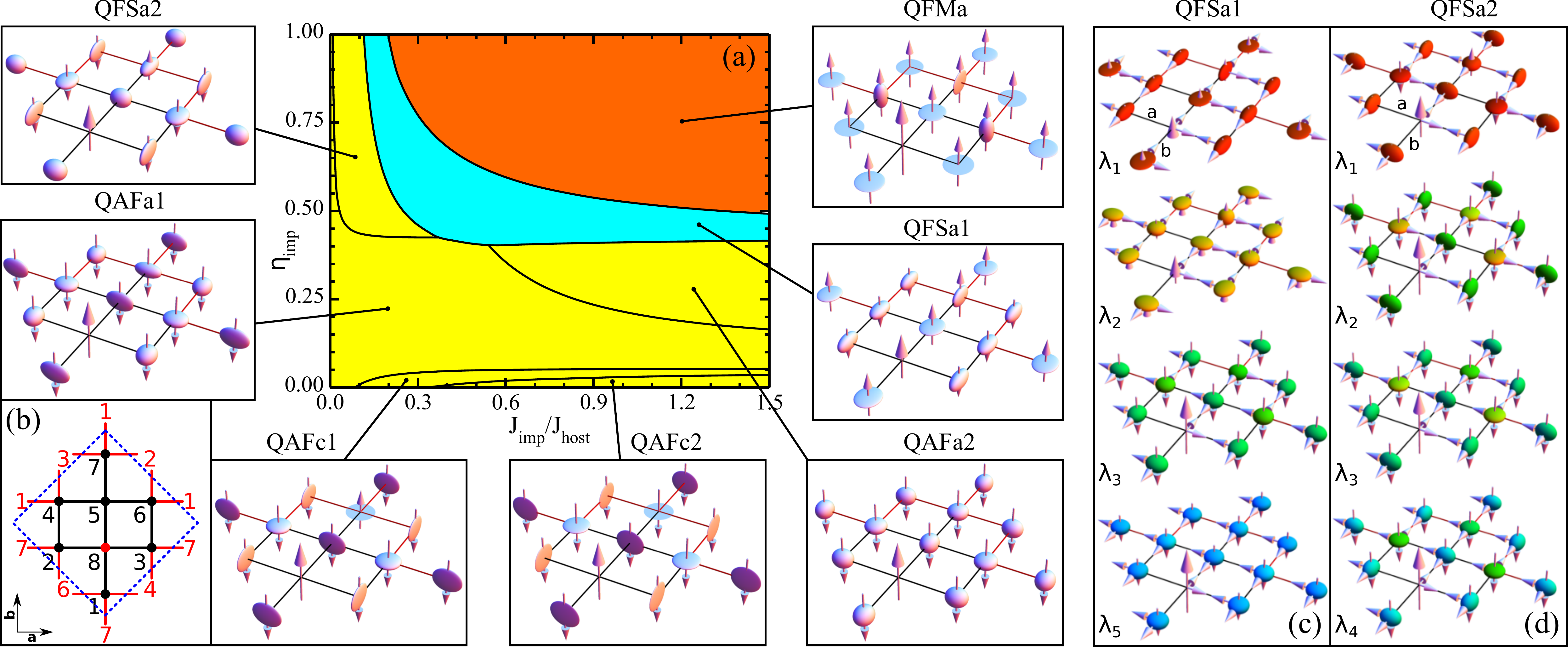}\end{center}\caption{Orbital dilution in the quantum limit. (a) Phase diagram as a function
of $J_{{\rm imp}}/J_{{\rm host}}$ and $\eta_{{\rm imp}}$ obtained
from a cluster of $8$ sites (b) with PBCs, without SOC. Arrows symbolize
values of local magnetization $\langle S_{i}^{z}\rangle$ and ellipsoids
local orbital polarization \textendash{} flat disk lying in the plane
means doublon in then $c$ orbital, disk perpendicular to $a$ and
$b$ direction means doublon in orbitals $a$ and $b$, spherical
ellipsoid means that all three orbitals are equally occupied, so lack
of polarization. On the right, (c)-(d) evolution of states QFSa1 and
QFSa2 as function of increasing spin-orbit coupling $\lambda_{p}$
on host sites. Vertical arrows are $\langle S_{i}^{z}\rangle$, in-plane
arrows symbolize sign of correlations $\langle S_{i}^{x}S_{j}^{x}\rangle$
(arrows in $a$ direction) and $\langle S_{i}^{y}S_{j}^{y}\rangle$
(arrows in $b$ direction), color of ellipsoids indicates average
value of local SOC term $\langle\vec{L}_{i}\vec{S}_{i}\rangle$. Reproduced
with permission from \cite{Brz15X}. Copyright 2015 American Physical
Society. \label{fig:8}}
\end{figure*}

\section{Orbital dilution in the quantum limit\label{sec:Orbital-dilution-in}}

Orbital dilution is even more interesting if we consider it in the
presence of quantum fluctuation \cite{Brz15X}. The full quantum phase
diagram for $x=1/8$ doping can be obtained by exact diagonalization
of an $8$ - site cluster with one impurity and PBCs, see Fig. \ref{fig:8}(a-b)
and evolution of different phases in presence of SOC at $4d^{4}$
host sites, shown in Fig. \ref{fig:8}(c). In case of a diagram \ref{fig:8}(a)
this coupling is absent so the total magnetization $M=\sum_{i}\langle S_{i}^{z}\rangle$
is a good quantum number. Other good quantum number is total number
of doublons $N_{a,b,c}$ in orbitals $a$, $b$ i $c$ of the host.
Different phases of the diagram \ref{fig:8}(a) have thus well defined
$M$ and $\{N_{a},N_{b},N_{c}\}$ and boundaries between the phases
are determined as level crossings of the lowest energy levels. The
representative spin-orbital configurations in the phases of the diagram
are shown in Fig. \ref{fig:8}(a); arrows stand for local magnetization
$\langle S_{i}^{z}\rangle$ and ellipsoids average occupation of orbitals
$a,$ $b$ and $c$ by a doublon, so the arrow without an ellipsoid
stands for impurity. The length of the semiaxes of the ellipsoids
in directions $a,$ $b$ and $c$ encodes the occupation of orbitals
$a$, $b$ and $c$ in such a way that if a semiaxis in direction
$\gamma$ is zero then the orbital $\gamma$ is occupied. For example,
if an ellipsoid looks like a disk in the plane of the cluster then
a doublon is almost exclusively in orbital $c$. On the other hand
a nearly spherical ellipsoid means that doublon does not favor any
orbital. There is some similarity of configurations shown in Fig.
\ref{fig:8}(a) to the phases of classical phase diagram \ref{fig:7}(c),
e.g., arrangement of orbitals around the impurity in phases QAFc1
and QAFc2 is similar to the one found in the AF' phase. A general
conclusion that can be drawn from the diagram \ref{fig:8}(a) is that
it is qualitatively similar to the diagram that can be obtain in classical
limit for small but finite concentration of impurities \cite{Brz15X},
and quantum fluctuations are most significant in FS phases where in
absence of quantum fluctuations remove frustration and are responsible
for polarization of the impurity spin. 

The evolution of representative configurations of the phase diagram
\ref{fig:8}(a) can be also traced for increasing values of spin-orbit
coupling $\lambda$ on host atoms. Such an evolution for two FS phases
is shown in Fig. \ref{fig:8}(c-d). The color of the ellipsoids means
local value of the SOC term $\langle\vec{L}_{i}\vec{S}_{i}\rangle$,
where a shift from red to violet means increase of this value. For
both phases such values of $\lambda=\lambda_{p}$ (increasing with
$p$) were chosen for which there is a significant change of spin
or orbital configurations. Due to the presence of SOC quantum numbers
$M$ and $N_{a,b,c}$ are no longer conserved so distinction between
different phases of diagram \ref{fig:8}(a) becomes collusive. The
universal behavior for large $\lambda$ are spin-orbital singlets
on every host site and residual magnetic moment at the impurity. An
interesting observation is that local $\langle\vec{L}_{i}\vec{S}_{i}\rangle$
for intermediate $\lambda$ has a non-trivial spatial distribution,
e.g., in phase QFSa2 sites $4$ and $6$ have much smaller $\langle\vec{L}_{i}\vec{S}_{i}\rangle$
than other sites. Another interesting effect, visible e.g. in phase
QFSa2 for $\lambda=\lambda_{2}$, is that for some bonds spin correlations
along $z$ axis (vertical arrows) have a different sign than along
the axes $x$ and $y$ (arrows in the cluster plane). This means that
in an analogical system but with spontaneous symmetry breaking the
spin order would be noncollinear, similarly as it happened for the
Kugel-Khomskii model for $d^{9}$ metals \cite{Brz12,Brz13}. However,
here the mechanism is more conventional \textendash{} SU$(2)$ symmetry
for spins is explicitly broken by the SOC terms in the Hamiltonian.

\begin{figure*}[t]
\begin{center}\includegraphics[width=1\textwidth]{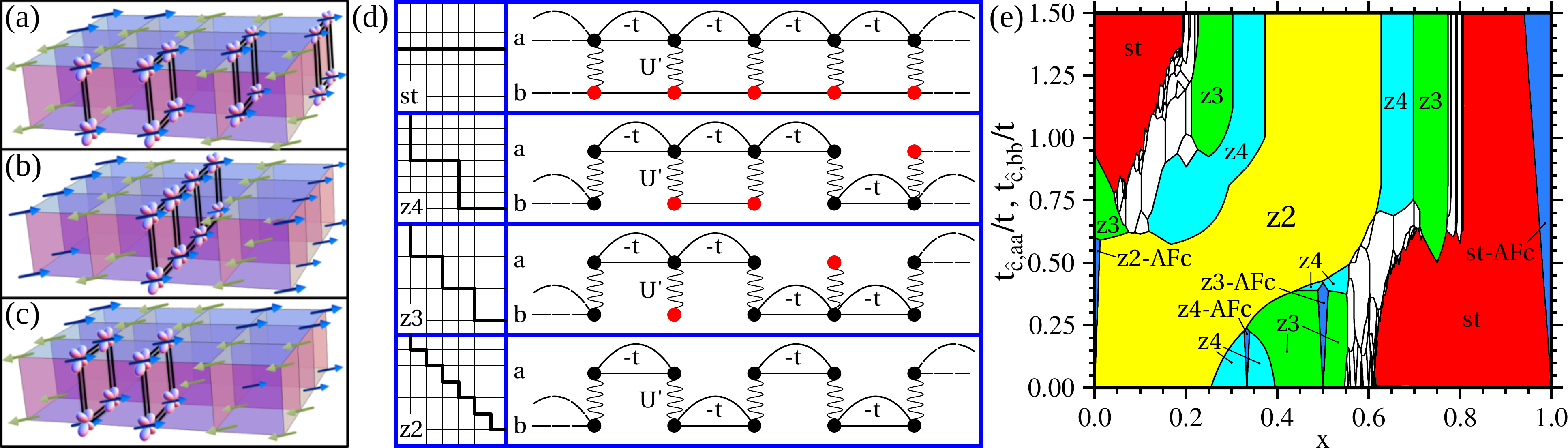}\end{center}\caption{(a-c) Schematic view of the magnetic phases for bilayer $d^{4}$ system
doped with $d^{3}$ ions obtained from the double-exchange mechanism;
(a) zigzag $z2$, (b) stripe phase $st$ and (c) checkerboard phase
$c2$. Double line means bonds for which hopping is possible within
a single magnetic domain and for orbitals which are shown. (d) 1D
magnetic domains in zigzag phases $zn$ and in stripe phase $st$
mapped on ladder systems; legs are the orbitals states $a$ and $b$,
sites marked in black are those for which hopping is possible with
amplitude $-t$ (arcs), in red those for which there is no in-plane
hopping. Wavy line marks interorbital Coulomb interaction $U'$. (e)
Phase diagram for 1D $zn$ and $st$ phases as a function of doping
$x$ and interplanar hopping amplitude $t_{\hat{c},aa}=t_{\hat{c},bb}$
(with respect to in-plane hopping $t$) in cases of $J_{H}\to\infty$,
i.e., when there is no hopping between opposite magnetic domains.
\label{fig:9}}
\end{figure*}

\section{Magnetic zigzag phases in the double-exchange model\label{sec:Magnetic-zigzag-phases}}

The problem of possible spin-orbital orders in the hybrid TMOs is
non-trivial not only in the limit of correlated insulator, where effectively
the charge degrees of freedom are absent, but also in the context
of a so-called double-exchange mechanism \cite{Zen51,Yun98}. Double-exchange
Hamiltonian we get by assuming that some of the charges get localized
giving rise to a magnetic order and by the Hund's exchange it affects
the energy of delocalized electrons described by kinetic Hamiltonian
$H_{t}$ (\ref{eq:Ht}). Such situation can arise due to so called
orbital selective Mott transition \cite{Anisimov2002,Bier05} when
electrons localize due to interactions only on some orbitals. A model
of this type was studied in Ref. \cite{Brz15L} in the context of
bilayer ruthenium $4d^{4}$ oxides doped with manganese $3d^{3}$
impurities. This is exactly the case which was called orbital dilution
in the limit of an insulator. The aim of this study was, \textit{inter alia},
explanation of experimentally observed magnetic phase with zigzag
order \cite{Mes12}, where in the plane spins order parallel along
zigzag lines and this pattern repeats in the next plane below, as
shown in Fig. \ref{fig:9}(a). The stability of various zigzag phases
was shown in the parameter range where FM and AF correlations compete
with each other \cite{Brz15L}. What more, these phases remain stable
in presence of octahedral distortions and finite interorbital Coulomb
interactions $U'$. Very interesting feature of these phases is that
the mechanism of their stability is purely kinetic and follows from
the directionality of $t_{2g}$ orbitals \textendash{} thanks to zigzag
kinks electrons can enclose themselves in 'orbital molecules' and
lower their kinetic energy with respect to propagation along straight
lines.

The double-exchange model considered in Ref. \cite{Brz15L} involves
three $t_{2g}$ orbitals at every lattice site, where all host atoms
have four electrons while impurities have only three, as shown in
Fig. \ref{fig:3}(c). We assume that $xy$ orbitals (or the $c$ ones)
are always singly occupied and electrons that occupy them localize
and order magnetically, whereas the electrons occupying orbitals $yz$
i $zx$ (or $b$ and $a$) can move freely according to Hamiltonian
$H_{t}$ (\ref{eq:Ht}). We assume that the ordering of the localized
spins $\vec{S}_{i}$, where $i$ labels lattice sites, is collinear
and purely classical, so that the Hund's interaction between localized
and itinerant electrons has a form of; $J_{H}S_{i}^{z}s_{i}^{z}$,
where $\vec{s}{}_{i}$ is the spins of these electrons. The choice
of the quatization axis as $z$ does not lower the generality of the
model if the order of spins $\{\vec{S}_{i}\}$ is collinear they do
not experience quantum fluctuations, the feature we assume here. The
interaction between localized spins $\vec{S}_{i}$ is then reduced
to; $J_{{\rm AF}}S_{i}^{z}S_{j}^{z}$, where $\{i,j\}$ are neighboring
lattice site (and $J_{{\rm AF}}$ is positive). The operator $s_{i}^{z}$
is a bilinear form of the creation $\{d_{i\gamma\sigma}^{\dagger}\}$
and annihilation $\{d_{i\gamma\sigma}\}$ operators of the itinerant
electrons in orbitals $\gamma=a,b$ and with spin $\sigma=\uparrow,\downarrow$.
Its form is given by; $s_{i}^{z}=\frac{1}{2}\sum_{\gamma=a,b}(d_{i\gamma\uparrow}^{\dagger}d_{i\gamma\uparrow}-d_{i\gamma\downarrow}^{\dagger}d_{i\gamma\downarrow})$.
Thus, for electrons $d_{i\gamma\sigma}^{\dagger}$ we get a quadratic
Hamiltonian parametrized by classical variables $S_{i}^{z}=\pm\frac{1}{2}$
living on every lattice site. Our task is to find such a configuration
of spins $S_{i}^{z}$ that for a given doping ratio $x$ (or electrons
$d_{i\gamma\sigma}^{\dagger}$) and the value of the coupling $J_{{\rm AF}}$
between spins $\{S_{i}^{z}\}$ gives minimal energy.

A similar optimization problem was solved by Dagotto and coauthors
for manganese oxides in Ref. \cite{Yun98}. The most general method
to solve it is to employ a classical Monte-Carlo simulation in variables
$S_{i}^{z}$. However, due to strong tendency of the system towards
phase separation a method of variational wave functions was used,
i.e., only some chosen ordered configurations of spins $S_{i}^{z}$
were considered and tested for the lowest energy for a given value
of $x$ and $J_{{\rm AF}}$ \cite{Brz15L}. These configurations are
either simple AF and FM phases or intermediate phases between these
two having a form of either AF zigzags with a segment length $n$,
labeled by $zn$, or straight stripes $st$ , or checkerboard phases
$cn$, with square $n\times n$ magnetic domains. Phases $z2$, $st$
and $c2$ are shown in Figs. \ref{fig:9}(a-c). It is not difficult
to predict that the FM phase will be stable for $J_{AF}\to0$, when
the system is dominated by kinetics, and the AF phase for $J_{{\rm AF}}\to\infty$,
when exchange interaction between localized spins is more important.
A non-trivial question is which phase is realized between these two
extreme cases.

\begin{figure*}[t]
\begin{center}\includegraphics[width=1\textwidth]{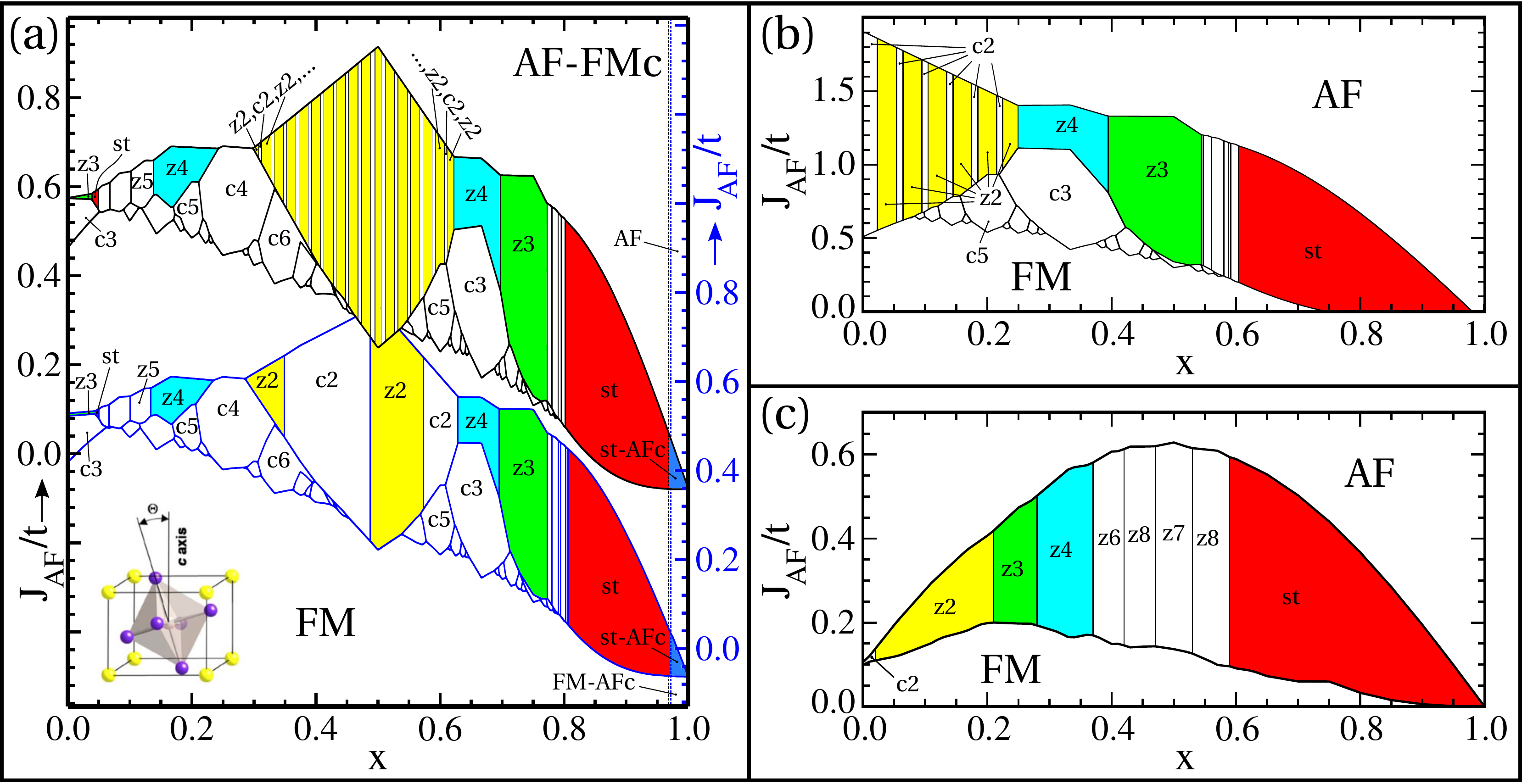}\end{center}\caption{Phase diagrams for the double-exchange model as functions of doping
$x$ and magnetic exchange $J_{{\rm AF}}$ for finite value of $J_{H}=100t$.
(a) Diagrams for $t_{\hat{c},aa(bb)}=0.8t$ for a system without distortions
(top plot and $J_{{\rm AF}}/t$ scale on the left) and with a tilting
octahedral distortion (shown in the inset) with a small angle $\theta=10^{\circ}$
(bottom plot and $J_{{\rm AF}}/t$ scale on the right). (b) Phase
diagram with no distortions for a single plane, i.e., $t_{\hat{c}}=0$
and (c) \textendash{} analogical diagram with interorbital Coulomb
interaction $U'=10t$. Phases with no labels are zigzags $zn$ or
checkerboards $cn$ with high $n$. \label{fig:10}}
\end{figure*}

An important factor affecting the stability of phase with 1D character
is directionality of orbitals; hopping of an electron in direction
$\hat{a}$(\foreignlanguage{english}{$\hat{b}$}) is possible only
through orbitals $b$($a$). For large $J_{H}$ with respect to $t$
(and this is a case of interest here) electrons with a fixed spin
$s_{i}^{z}$ are almost exclusively in the domains with the same spin
$S_{i}^{z}$, because hopping to the opposite domain costs energy
$J_{H}$. Thus, in the limit of $J_{H}\to\infty$ we can treat zigzag
phases a set of independent 1D subsystems and a single zigzag or a
stripe map on a ladder, depicted in Fig. \ref{fig:9}(d). The legs
of this ladder are orbitals $a$ and $b$ in such a way that every
rung is a single site of a 2D lattice. For a phase $st$ with stripes
in $\hat{b}$ direction hopping on a ladder are possible only along
leg $a$. Bending the stripes and forming zigzags we change orbitals
through which we can hop, as shown in Fig. \ref{fig:9}(d). In case
of the shortest zigzag $z2$ for every magnetic domain we effectively
get a set of independent two-site molecules formed at neighboring
lattice sites with well defined orbital polarization \textendash{}
hence this is a phase with orbital order. Fig. \ref{fig:9}(e) shows
a phase diagram for such 1D phases where the parameters are doping
$x$ and interplane hopping amplitude $t_{\hat{c}}$. The diagram
shows that $z2$ phase is stable in quite wide range of doping around
$x=0.5$ in case when amplitude $t_{\hat{c}}$ is equal to inplane
hopping amplitude~$t$. This interval shifts towards $x=0$ if we
decrease $t_{\hat{c}}$ and for $t_{\hat{c}}=0$ (independent planes)
the $z2$ phase starts from zero doping, meaning high electron density.
This is related with a competition between inplane and interplane
orbital molecules. A consequence of having second plane is also presence
of phases with AF magnetic correlations in $\hat{c}$ direction, such
as $st$-AFc phase, which however occur only in narrow windows of
doping.

The above considerations concern only 1D phases, i.e., zigzags and
stripes. A generic 2D case is richer because it involves hopping processes
between magnetic domains, always present for finite $J_{H}$, and
phases with a 2D character, like AF, FM and $cn$ ones. Additionally,
it is possible to describe a system with octahedral distortions that
allow for hybridization of orbitals on the bonds. In Ref. \cite{Brz15L}
two types of distortions were considered; cooperative rotation of
the oxygen octahedra and tilting of the octahedra with respect to
axis perpendicular to the plane. The first case turns out to be trivial
\textendash{} the system with rotation distortion in this case is
equivalent to the system with no distortion. On the other hand, the
tilting distortion is non-trivial and affects phase diagrams of the
system, which is shown in Fig. \ref{fig:10}(a) with phase diagrams
for tilting angle $\theta=0$ and $\theta=10^{\circ}$ obtained for
$t_{\hat{c},aa(bb)}=0.8t$ and relatively large $J_{H}=100t$. In
these diagrams we see a window bewteen FM and AF phases where zigzag
$zn$ and stripe $st$ phases are formed. However, these phases have
to compete with checkerboard $cn$ phases which grow from below as
a extension of the FM phase. Because of similar band structures the
competition (almost degeneracy) between $z2$ and $c2$ phases is
particularly strong in the doping region around $x=0.5$ but only
for $\theta=0$ and the effect vanishes in presence of distortions.

In case of a single plane, i.e., $t_{\hat{c},aa(bb)}=0$, the diagram
without distortions is shown in Fig. \ref{fig:10}(b). Zigzag phase
$z2$ shifts towards zero doping, as in the 1D diagram \ref{fig:9}(e),
but as it was before still strongly compete with checkerboard $c2$.
In this case however there it was possible to find another mechanism
increasing the stability of the $z2$ phase \textendash{} interorbital
Coulomb interaction $U'$ (according to the general Hamiltonian (\ref{eq:Hint}),
$U'=U-\frac{5}{2}J_{H}$), such as schematically depicted in Fig.
\ref{fig:9}(d). Diagram for a single plane with interactions $U'=10t$
is shown in Fig. \ref{fig:10}(c). In presence of interaction between
electrons $d_{i\gamma\sigma}^{\dagger}$ the model becomes a many-body
problem which is unsolvable, so the results were obtained via exact
diagonalization of finite systems assuming $J_{H}\to\infty$, meaning
that the electrons do not leave their magnetic domains.Under such
assumption it is correct to include only interorbital interactions
due to the absence of electrons with opposite spins. As one can see
from diagram \ref{fig:10}(c), checkerboard phases do not appear almost
at all and the order of the zigzag phases in the window between FM
and AF regions is similar to the one found in the former case without
interactions. An interesting feature is that, apart from quite obvious
magnetic and orbital orders realized by the zigzag configurations,
in the $z3$ phase a non-trivial charge order was found, i.e., the
optimal charge distribution is alternating between $2$ and $1$ electrons
in every odd/even segments of the zigzag. This means that phase $z3$
can have a non-vanishing electric polarization or ferroelectric order.
Another effect, common for all diagrams \ref{fig:10}(a-c) is exotic
metal-insulator transition between phase $st$, being metallic, and
phase $z2$ where electrons are fully localized. It takes place by
a cascade of zigzag phases whose segment length diverges when we approach
the $st$ phase. This is quite peculiar transition from a molecular
insulator to a 1D metal by the growth of the molecules, or zigzag
segments.

Summarizing, in was shown in Ref. \cite{Brz15L} how interestingly
a spin-orbital-charge order of the $d^{4}$ host can change in presence
of $d^{3}$ doping in the limit of partially localized electrons,
i.e., localized but only for one orbital flavor. In this case the
spin-charge density modulations is due to the purely kinetic mechanism
of the $t_{2g}$ electrons but a similar type of order can arise from
spin-orbital superexchanges in models of insulating $t_{2g}$ electrons
\cite{Wro10,Tro13}. Such exotic spin orders provoke a natural question
of propagation of charge in such systems, a property that can be seen
by a photoemission experiment, which is especially challenging in
fully insulating regiome \cite{Dag08,Tro13,Brz14-adhoc,Brz15-adhoc}.
On the other hand, in the double exchange limit zig-zag orderings
open a route towards exotic topological semi-metal \cite{Brz17B}
or nodal superconductor \cite{Brz18-nodal} phases, extablishing a
connection between magnetism and topological matter.

\begin{figure*}[t]
\begin{center}\includegraphics[width=0.7\textwidth]{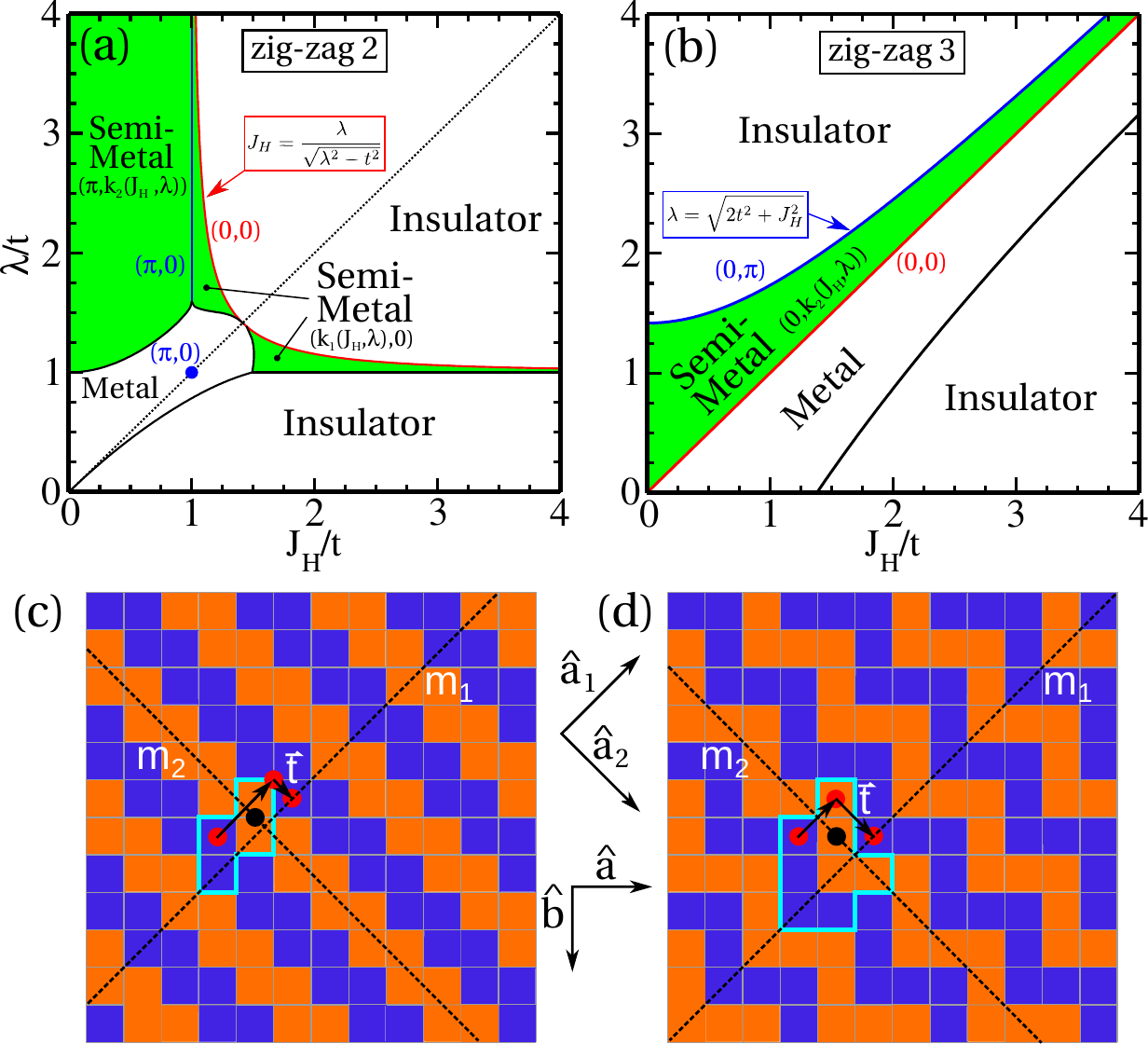}\end{center}\caption{Electronic phase diagrams for systems with zigzag magnetic orders
and their schematic view. (a) and (b) Diagrams for $z2$ system with
$3/4$ filling and $z3$ with $1/2$ filling, green regions are topological
semi-metal phases with Dirac points. (c) and (d) View of $z2$ and
$z3$ magnetic orders, color regions are magnetic domains with spins
up (orange) and down (blue), small squares symbolize lattice sites.
Dashed lines are the mirror reflection lines $m_{1}$ and $m_{2}$,
black dots are inversion centers and vector $\vec{t}$ is a shift
that together with reflection $m_{2}$ forms a nonsymmorphic symmetry
\textendash{} a glide. Action of this symmetry is depicted with red
dots. Elementary cells are marked with light blue frames\label{fig:11}}
\end{figure*}

\section{Topological semi-metal phases in systems with zigzag magnetic order\label{sec:Topological-semi-metal-phases}}

The topological issues of itinerant electrons in the presence of the
zigzag magnetic order were exensively studied in Ref. \cite{Brz17B}
for $z2$ and $z3$ orders, whose stability in doped $d^{4}$ oxides
was addressed earlier \cite{Brz15L}. These configurations were found
to have topological semi-metal phases with Dirac points (DPs) as functions
of Hund's exchange and SOC. In case of zigzag $z2$ these points have
coexisting topological charges of different types and this follows
from simultaneous presence of many symmetries in the system, including
a nonsymmorphic symmetry \textendash{} in this case a mirror reflection
with a shift of half lattice translation. In case of zigzag $z3$
this symmetry leads, together with another mechanism described in
Ref. \cite{Brz15L}, to double DPs, i.e., a linear band touching with
degeneracy $d=4$. This is local dispersion of a relativistic particle
with spin $S=3/2$. Every time the presence of topological charges
manifests itself by a presence of topologically protected edge states.
Thus we see that the coexistence of magnetic order and nonsymmorphic
symmetries can lead to exotic topological properties.

The main focus of study in Ref. \cite{Brz17B} is determining the
symmetries behind the topological protection of DPs and the behavior
of DPs if one breaks them. The mechanism of the mirror symmetry protection
is known and there exist tables with the classification of topological
invariants that one can have depending on the Altland-Zirnbauer class
of the Hamiltonian, spatial dimension and commutation relation between
reflection operator and time-reversal and particle-hole symmetries
\cite{Chi14}. Similar classification tables exist for topological
states protected by nonsymmorphic symmetries in gapped systems \cite{Shi16}
but not in the gapless cases, so the question of topological properties
of nonsymmorphic gapless systems is still valid \cite{Kob16,Brz17B}.

In the zigzag model considered in Ref. \cite{Brz17B} it is assumed
that electrons in a 2D system hop through orbitals $xz$ and $yz$
and experience Hund's interaction $J_{H}S_{i}^{z}s_{i}^{z}$ with
localized spins $S_{i}^{z}$ at orbitals $xy$. Additionally, they
are subject to anisotropic spin-orbit interaction $\lambda S_{i}^{z}l_{i}^{z}$,
where $l_{i}^{z}=i(d_{i,a,\sigma}^{\dagger}d_{i,b,\sigma}-d_{i,b,\sigma}^{\dagger}d_{i,a,\sigma})$,
being projection of the full interactions term on the subspace of
orbitals $xz$($a$) and $yz$($b$). Both interaction conserve spin
$\sigma$ of itinerant electrons $d_{i,\gamma,\sigma}^{\dagger}$,
which is a good quantum number. Effectively, the problem is reduced
to a Hamiltonian of free and spinless fermions where Hund's interaction
$J_{H}$ enters as a chemical potential spatially modulated by $S_{i}^{z}$
and spin-orbit coupling $\lambda$ as an on-site hopping between orbitals
$a$ and $b$ with its amplitude modulated by $S_{i}^{z}$.

For zigzag systems $z2$ and $z3$ electronic phase diagrams were
determined as functions of $J_{H}$ and $\lambda$, shown in Figs.
\ref{fig:11}(a) and \ref{fig:11}(b). It turned out that topological
states (green areas in the diagrams) can be observed for the $z2$
system with $3/4$ filling and $z3$ one for half-filling and they
appear in a semi-metal phase, where energy gap closes at isolated
points in the momentum space \textendash{} Dirac points. These points
are placed at high-symmetry lines of the BZ; their positions are marked
in diagrams as $\vec{k}=(k_{1},k_{2})$. In case of zigzag $z2$ this
is either $k_{1}=\pi$ line (main part of the topological phase for
$J_{H}<t$) or $k_{2}=0$ one (smaller parts adjacent to insulator
phase). On the other hand, for zigzag $z3$ DPs are always in the
line $k_{1}=0$. Values of the remaining components of $\vec{k}$
are functions or parameters $J_{H}$ i $\lambda$. For example, in
the $z2$ system being in the main semi-metal phase and increasing
$\lambda$ we shift DPs along the line $k_{1}=\pi$ until they merge
at high-symmetry point $\vec{k}=(\pi,0)$ for $\lambda=t$ and, further
increasing $\lambda$, they split again but now in the $k_{2}=0$
line. Further increase of $\lambda$ makes the point merge in another
high-symmetry point $\vec{k}=(0,0)$ and then energy gap occurs and
the system undergoes transition to an insulating state. Similarly,
in the $z3$ system, being in the diagonal of phase diagram ($J_{H}=\lambda$)
and moving perpendicularly to it through semi-metal phase we change
$k_{2}$ from $0$ to $\pi$ keeping constant $k_{1}=0$. This means
that DPs move from high symmetry point $\vec{k}=(0,0)$ to $\vec{k}=(0,\pi)$
along the line $k_{1}=0$. 

\begin{figure*}[t]
\begin{center}\includegraphics[width=1\textwidth]{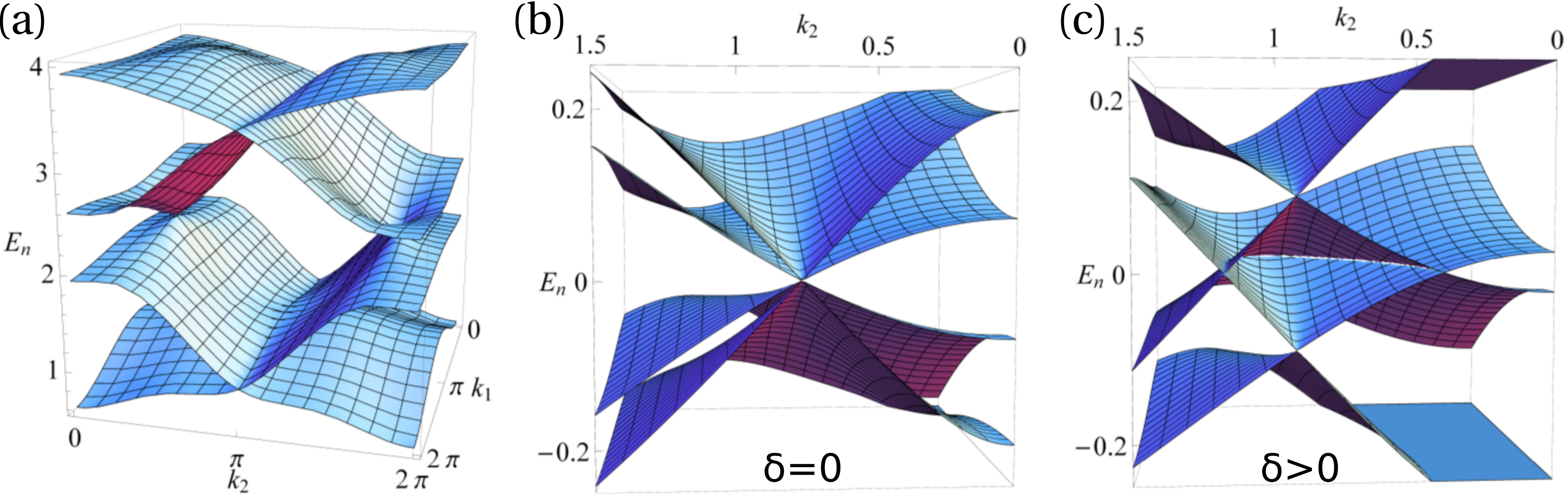}\end{center}\caption{Energy bands in semi-metal phases. (a) $z2$ system with Dirac points
in the glide line $k_{1}=\pi$ for $3/4$ filling (only positive energy
bands are shown, bands are symmetric with respect to zero energy).
(b) $z3$ system with a double Dirac point, also in the glide line,
$k_{1}=0$. This point evolves into a Fermi circle (c) or a nodal
line for a finite $\delta$, extra hopping term in the Hamiltonian,
that preserves all symmetries of the system. \label{fig:12}}
\end{figure*}

Schematic views of the $z2$ and $z3$ systems are presented in Figs.
\ref{fig:11}(c) and \ref{fig:11}(d), which also shows elementary
cells, related lattice directions $\hat{a}_{1}$ and $\hat{a}_{2}$
(corresponding quasimomenta are $k_{1}$ and $k_{2}$) and spatial
symmetries. These symmetries are; mirror reflection with respect to
$m_{1}$, mirror reflection with translation $\vec{t}$ and spatial
inversion symmetry. Reflection with translation is called a glide
and it is a nonsymmorphic symmetry, i.e., such that is composed of
point group symmetry and translation by vector which is a fraction
of a lattice vector. In our case for both systems $\vec{t}=\hat{a}_{2}/2$
and this is a parallel direction to reflection line $m_{2}$. Action
of this symmetry on a single lattice site is demonstrated in Fig.
\ref{fig:11}(c) and \ref{fig:11}(d). These figures do not show however
orbital degrees of freedom \textendash{} since hopping in direction
$\hat{a}$($\hat{b}$) is possible only through orbitals $b$($a$),
and mirror reflections $m_{1,2}$ interchange $\hat{a}$ and $\hat{b}$,
it is necessary to interchange orbitals $a$ and $b$ within reflection
operator as well. Another type of symmetry is a sublattice symmetry
or chirality ${\cal S}$ being an interchange of magnetic domains
within a unit cell. This can be achieved by a translation by a vector
$\vec{s}=\hat{a}_{1}/2$. Since this is a fraction of a lattice translation,
chirality ${\cal S}$ is also a nonsymmorphic symmetry.

\begin{figure*}[t]
\begin{center}\includegraphics[width=1\textwidth]{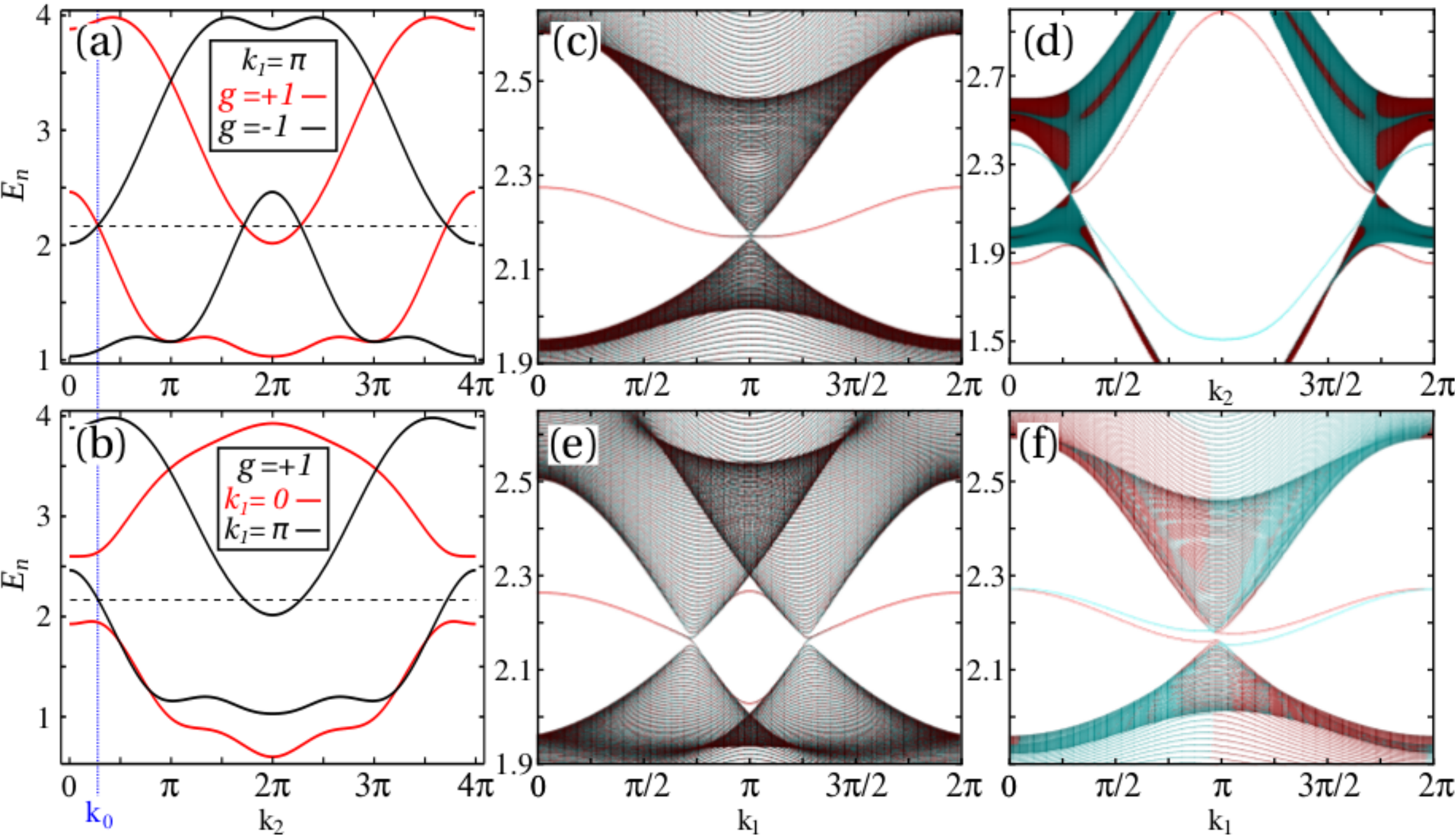}\end{center}\caption{Energy bands and edge states for a $z2$ system. (a) Bands for $k_{1}=\pi$
with fixed eigenvalues of the glide operator; ${\cal R}^{t}=+1$ for
red lines and ${\cal R}^{t}=-1$ for black ones, $k_{2}=k_{0}$ marks
position of the DPs. (b) Bands for ${\cal R}^{t}=+1$ and two glide
planes; $k_{1}=0$ for red lines and $k_{1}=\pi$ for black ones.
Dashed lines in plots (a-b) is a Fermi energy at the DPs. (c) and
(d) Energy spectra for half-open systems in $\hat{a}_{2}$ and $\hat{a}_{1}$
directions, where every dot is a state with fixed energy and quasimomentum
in direction parallel to the system's edge. (e) and (f) Energy spectra
for half-open systems in $\hat{a}_{2}$ directions with broken symmetry
of; (e) glide but preserved inversion and (f) time reversal. Brightness
of the red/green color of a dot means localization of a state on left/right
edge of the system. Reproduced with permission from \cite{Brz17B}.
Copyright 2017 American Physical Society.\label{fig:13}}
\end{figure*}

In the momentum space Hamiltonians ${\cal H}_{\vec{k}}$ for $z2$
and $z3$ systems are represented by matrices of sizes $8\times8$
and $16\times16$, respectively. The symmetry operators are represented
in a similar way; reflection ${\cal R}$ (with respect to $m_{1}$)
and glide ${\cal R}^{t}$ (with respect to $m_{2}$), where the latter
one involves intrinsic dependence on $k_{2}$ due to translation $\vec{t}$.
Action of these operators on Hamiltonian is to reverse the sign of
quasimomentum perpendicular to the line of reflection, i.e., ${\cal R}^{\dagger}{\cal H}_{k_{1},k_{2}}{\cal R}={\cal H}_{k_{1},-k_{2}}$
and ${\cal R}^{t\dagger}{\cal H}_{k_{1},k_{2}}{\cal R}^{t}={\cal H}_{-k_{1},k_{2}}$.
Lines of high symmetry $k_{1(2)}=0,\pi$ correspond to the reflection
lines $m_{2(1)}$ and in these parts of BZ operators ${\cal R}^{t}$(${\cal R}$)
commute with Hamiltonian ${\cal H}_{\vec{k}}$, meaning that energy
bands $E_{n}(\vec{k})$ can be labeled by quantum numbers being the
eigenvalues of ${\cal R}^{t}$(${\cal R}$). On the other hand, chiral
symmetry ${\cal S}$ for half-filled systems satisfies ${\cal S}^{\dagger}{\cal H}_{\vec{k}}{\cal S}=-{\cal H}_{\vec{k}}$,
so it anticommutes with Hamiltonian for any $\vec{k}$ and analogically
to ${\cal R}^{t}$ contains intrinsic dependence on $k_{1}$ due to
translation $\vec{s}$. Intrinsic dependence on quasimomentum implies
that one cannot define such a unit cell that would map onto itself
under the action of operators ${\cal R}^{t}$ or ${\cal S}$ \cite{Brz17B}.

The plots of the energy bands with DPs for $z2$ and $z3$ systems
are shown in Figs. \ref{fig:12}(a) and \ref{fig:12}(b). A very interesting
feature for $z3$ system is that its DPs have fourfold degeneracy
and energy bands around these points have form of four Dirac cones
touching with their tips. These cones can be split without breaking
any symmetry of the system by adding to the Hamiltonian extra hopping
terms $\delta$ \cite{Brz17B}, and then one obtains Fermi surfaces
of the form of doubly degenerate circles or nodal lines, as shown
in Fig. \ref{fig:12}(c). On the other hand, DPs of the $z2$ zigzag
have more conventional form and resemble DPs found in graphene \cite{Net09,Ryc07}. 

In case do DPs lying in the glide line one can expect that this is
the symmetry that protects them from hybridization and opening a gap.
In Fig. \ref{fig:13}(a) one can see that this is indeed true in case
of $z2$ zigzag; it shows energy bands for $k_{1}=\pi$ whose color
correspond to two eigenvalues of ${\cal R}^{t}$. It is worth to notice
that such colored bands have period of $4\pi$, not $2\pi$, which
follow from the nonsymmorphic character of symmetry ${\cal R}^{t}$
\cite{Brz17B}. DPs in Fig. \ref{fig:13}(a) appear as crossing points
of bands with different eigenvalues of ${\cal R}^{t}$, which indicates
that they are protected by the glide. As it follows from a general
theory \cite{Chi14}, DPs in such case have topological charge $M\mathbb{Z}$,
which manifests itsefl by the edge states in a system with open edge.
Such states can be seen as colored bands in Fig. \ref{fig:13}(c),
where the system is open in $\hat{a}_{2}$ direction but keeps translation
symmetry in $\hat{a}_{1}$ direction. These bands have double degeneracy
arising from the fact that one is located on the right and one on
the left edge of the system. They connect two DPs (in Fig. \ref{fig:13}(c)
they overlap) with opposite topological charges and in this case they
have an energy gap and finite dispersion. The lack of flat band, observed
for instance in graphene, results from the fact that the edge of system
is not invariant with respect to ${\cal R}^{t}$ \textendash{} it
is easy to check that in two dimensions there is no such edge.

Fig. \ref{fig:13}(d) shows spectrum for the $z2$ system open in
$\hat{a}_{1}$ direction, so a case complementary to that depicted
in \ref{fig:13}(d). Now we can see both DPs connected directly with
two non-degenerate edge states and with one through the boundary of
the BZ. This third edge state indicates additional topological charge
of the DPs. Its existence can be proven by looking at the bands in
1D subsystems with fixed $k_{2}$, perpendicular to the glide line,
shown in Fig. \ref{fig:13}(b). For each such subsystem ${\cal R}^{t}$
is an inversion symmetry operator and each one of them, except those
crossing DPs at $k_{2}=\pm k_{0}$, has energy gap. As it was proven
in Ref. \cite{Ale14} they can have a non-trivial topological index
$\mathbb{Z}^{\geq}$ related to the inversion symmetry, which can
be expressed as a difference in number of occupied states with fixed
inversion eigenvalue between high-symmetry points $k_{1}=0$ and $k_{1}=\pi$.
Fig. \ref{fig:13}(c) shows such states as functions of $k_{2}$,
where different colors of bands correspond to points $k_{1}=0$ and
$k_{1}=\pi$. One can see that for $k_{2}<k_{0}$ the value of $\mathbb{Z}^{\geq}$
index is non-vanishing because the number of red and black lines below
the Fermi level is different. For $k_{2}=k_{0}$ there is a topological
phase transition to a trivial phase where $\mathbb{Z}^{\geq}$ vanishes.
DPs are thus boundaries between topologically trivial and non-trivial
phases for 1D subsystems perpendicular to the glide line. This is
consistent with appearance of an extra edge state in Fig. \ref{fig:13}(d)
for $k_{2}<k_{0}$ and $k_{2}>2\pi-k_{0}$.

\begin{figure*}[t]
\begin{center}\includegraphics[width=1\textwidth]{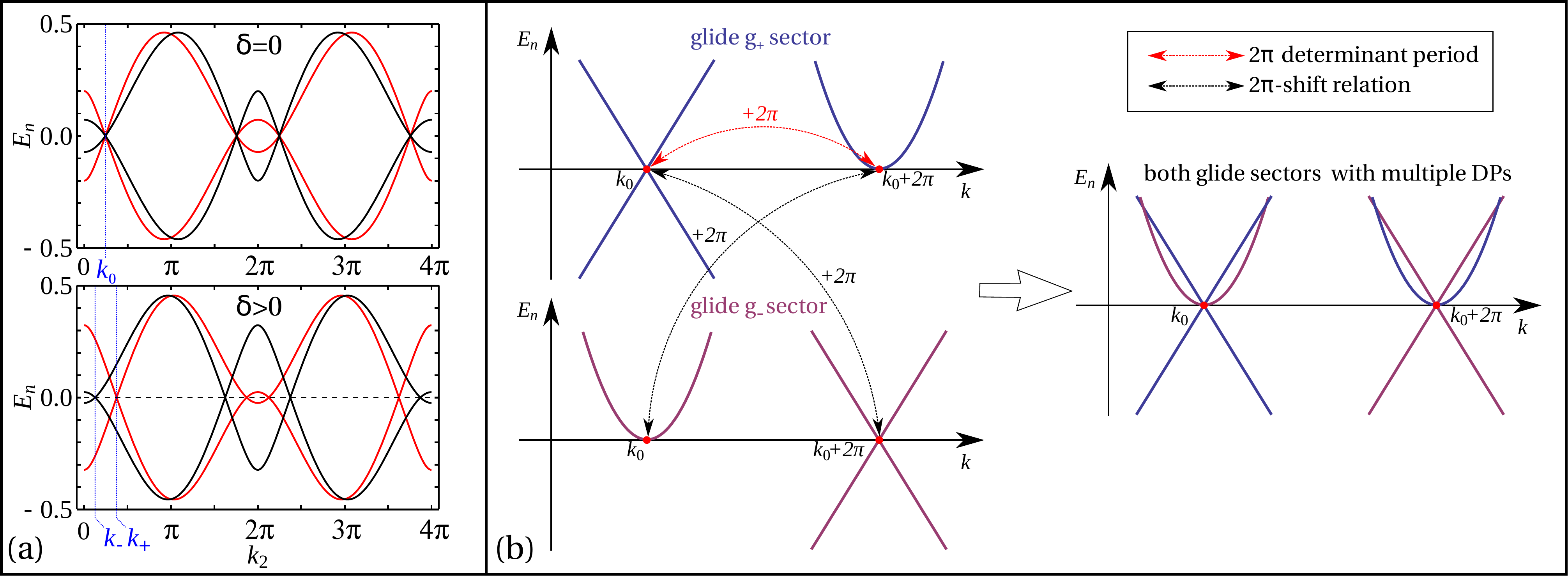}\end{center}\caption{(a) Energy bands for $k_{1}=0$ and zigzag $z3$ and (b) mechanism
of forming multiple DPs (or more general multiple Fermi points). Bands
have fixed values of glide; ${\cal R}^{t}=+1$ for red lines and ${\cal R}^{t}=-1$
for black ones, both zero and finite $\delta$ cases are shown. \label{fig:14}}
\end{figure*}

The last topological charge, which can be assigned to DPs of the $z2$
system is the $\mathbb{Z}_{2}$ index related to simultaneous presence
of inversion and time-reversal symmetries \cite{Zha16}. Its existence
can be determined by breaking both reflection and glide symmetries
but keeping their product which is inversion. It turns out that this
does not open a gap but only moves DPs outside the glide line $k_{1}=\pi$.
A spectrum for a half-open system with edge states is presented in
Fig. \ref{fig:13}(e). They could not exist without a third, $\mathbb{Z}_{2}$
topological charge at the DPs, which can survive breaking of the ${\cal R}^{t}$
symmetry. Another example of symmetry breaking is breaking of the
time-reversal. Fig. \ref{fig:13}(f) shows the spectrum of a half-open
system where it happened without breaking other symmetries. As one
can see an infinitesimal gap opens in the bulk and it is closed by
non-degenerate edge states connecting bottom bands with the upper
ones. The spectrum resembles a classical case of a 2D topological
insulator with a non-vanishing Chern number or a quantum Hall system.

These considerations concerned only magnetic phase $z2$. State of
a topological semi-metal for the $z3$ configuration is different
because of a chiral symmetry ${\cal S}$ present at half-filling,
that affects possible topological charges of the Fermi surface. It
turns out that double DPs place in the glide line $k_{1}=0$ can be
continuously and without breaking any symmetry of the system transformed
into Fermi circles crossing the $k_{1}=0$ line at four ordinary DPs,
as one can see in Fig. \ref{fig:14}(a). Such circles have topological
charge $\mathbb{Z}_{2}$ arising from simultaneous presence of inversion
and particle-hole symmetries \cite{Zha16}. The effect of transformation
of multiple DPs into Fermi circles or nodal lines was not described
in any earlier work concerning topological systems. We remark that
it was possible to describe a hidden non-unitary symmetry which allows
for existence of multiple DPs at $\delta=0$.

Characteristic feature of a nonsymmorphic symmetry such as ${\cal R}^{t}$
is that at the symmetry line, here $k_{1}=0$, bands with fixed eigenvalue
of ${\cal R}^{t}$ have a period not of $2\pi$ but of $4\pi$. On
the other hand, since the full Hamiltonian has a period of $2\pi$,
these bands cannot be independent \textendash{} they must differ at
most by a shift of $2\pi$. This indeed happens in Fig. \ref{fig:14}(a).
Additionally, in case of the $z3$ zigzag at $\delta=0$, it turns
out that despite bands having period of $4\pi$ the determinant of
the Hamiltonian in each eigenspace of ${\cal R}^{t}$ has still a
period of $2\pi$, so the bands are $4\pi$-periodic but their product
is already $2\pi$-periodic. This indicates that we have some symmetry
of the Hamiltonian at $\delta=0$ which however refers not to the
operator itself but to its determinant. For this reason it is non-unitary,
as it was shown in Ref, \cite{Brz17B}. Having the knowledge about
the determinant it is easy to explain the mechanism of forming multiple
DPs, or more generally, multiple bands touching points at the Fermi
level. This mechanism is depicted in Fig. \ref{fig:14}(b). In a subspace
of fixed value of glide (or other order two nonsymmorphic symmetry)
we have a Fermi point at $k=k_{0}$, where its origin and degeneracy
can be any. Due to the property of determinant it must repeat at $k=k_{0}+2\pi$
in the same subspace although its degeneracy and dispersion can be
different. In the other subspace the bands differ only by a shift
of $2\pi$ so $k_{0}$ and $k_{0}+2\pi$ are still Fermi points, but
interchanged. Now, taking both subspaces together we immediately see
that at $k_{0}$ and $k_{0}+2\pi$ we get multiple Fermi points.

The mechanism described above is an interesting peculiarity of a nonsymmorphic
symmetry. It is responsible not only for degeneracy $d=4$ DPs in
zigzag $z3$ at $\mu=0$ but also for DPs with $d=3$ for 'magical'
value of chemical potential $\mu=\mu_{0}=\sqrt{2t^{2}+J_{H}^{2}+\lambda^{2}}$.
Such a triple Dirac point is also found in the glide line $k_{1}=0$
and it consists of an ordinary DP crossed by a weakly dispersive parabolic
band that contributes to the Fermi surface at $\mu_{0}$. This case
resembles Dirac points crossed by flat bands found in the Lieb lattice
models \cite{Wee10}, which can be relavant for TMOs, however here
the lattice structure is simpler and the effect is caused by a nonsymmorphic
magnetic pattern.

\section{Topological phases in non-uniform Kitaev model\label{sec:Topological-phases-in}}

As shown in Sec. \ref{sec:Inhomogeneous-spin-orbital-model} doping
of a $d^{4}$ host with a $d^{2}$ metal can lead in the superexchange
limit to a Hamiltonian with orbital terms $\tau_{i\gamma}^{+}\tau_{j\gamma}^{+}$
around the dopants, where $\vec{\tau}_{i\gamma}$ are orbital pseudospins
$\tau=1/2$ and $\tau_{i\gamma}^{\pm}$ are lowering/raising operators
of $\tau_{i\gamma}^{z}$. Apart from this, on every bond, both of
host and around impurities, there are also terms $\tau_{i\gamma}^{+}\tau_{j\gamma}^{-}$
and $\tau_{i\gamma}^{z}\tau_{j\gamma}^{z}$ . For a 1D system, for
instance along $\gamma=a$, we have effectively an $XXZ$ Heisenberg
model on host bonds and $XYZ$ on bonds around impurities. 

\begin{figure*}[t]
\begin{center}\includegraphics[width=0.8\textwidth]{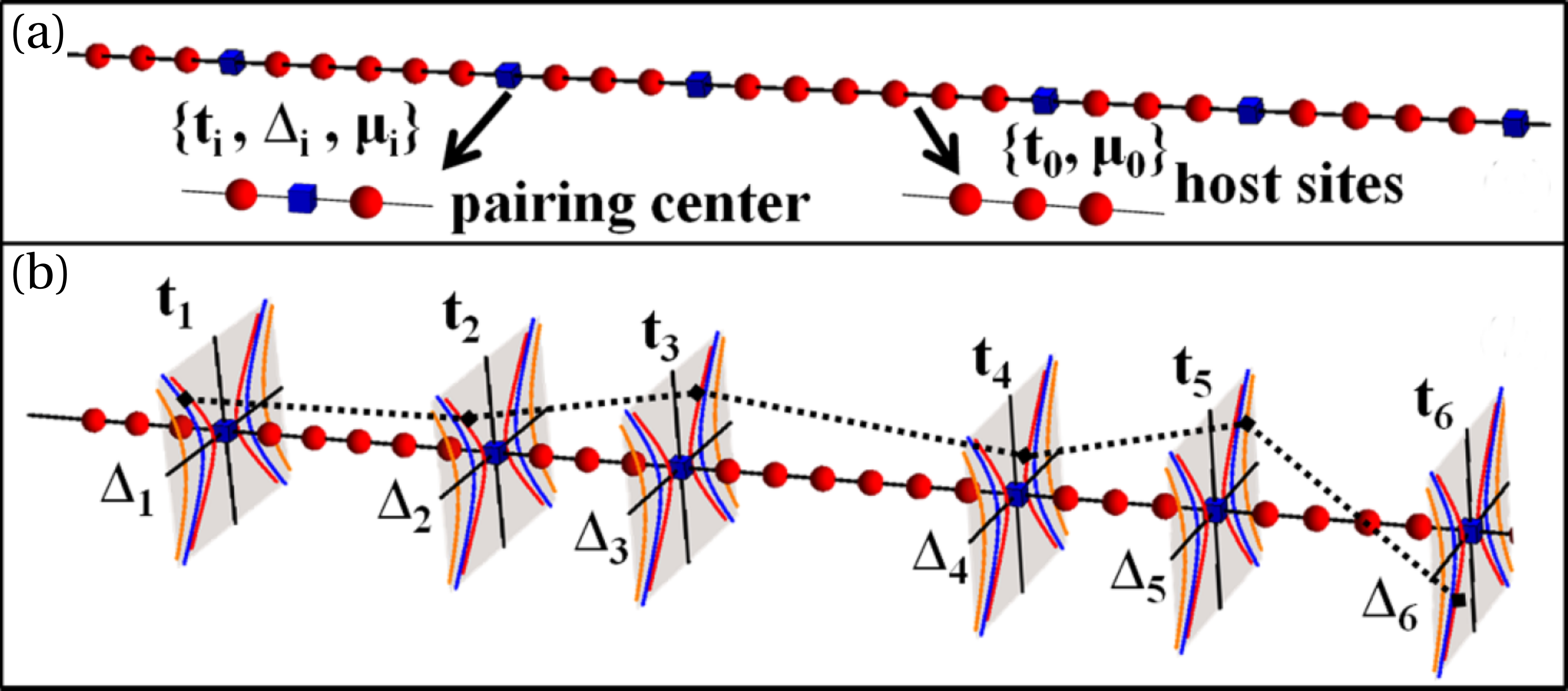}\end{center}\caption{Non-uniform 1D superconducting system effectively realized by a charge
dilution and orbital pairing mechanism. (a) Schematic view of the
system, red balls are $d^{4}$ host atoms and blue ones are $d^{2}$
impurities being pairing centers. Hopping amplitude between host atoms
is $t_{0}$. Between impurities and host sites there is hopping $t_{i}$
and pairing $\Delta_{i}$, where $i=1,\dots,N$ labels impurities.
Host and impurities atoms have chemical potentials $\mu_{0}$ and
$\mu_{i}$. (b) Artist's view of the hidden Lorentz symmetry of the
topological phase, this phase is stable with respect to change of
parameters $t_{i}$ and $\Delta_{i}$ at every impurity site under
condition that $r_{i}^{2}=t_{i}^{2}-\Delta_{i}^{2}$ is left unchanged,
so that these parameters are lying on hyperbolas with $r_{i}$ radii.
Reproduced with permission from \cite{Brz17R}. Copyright 2017 American
Physical Society. \label{fig:15}}
\end{figure*}

In one dimension pseudospin operators $\vec{\tau}_{i}$ can be easily
mapped onto spinless fermions using Jordan-Wigner transformation.
In this way terms $\tau_{i}^{+}\tau_{i+1}^{-}$ become hopping between
nearest neighbors and $\tau_{i}^{+}\tau_{i+1}^{+}$ become paring
terms, as the ones in the $p$-type Kitaev superconductor \cite{Kit09}.
Furthermore, if we substitute the diagonal $\tau_{i}^{z}\tau_{i+1}^{z}$
terms by the mean-field $h_{i}\tau_{i}^{z}$ terms (from the point
of view of fermions this is Hartree decoupling) then we obtain local
chemical potential different for host and impurity sites. Thus, we
obtain a non-uniform superconducting Kitaev model schematically depicted
in Fig. \ref{fig:15}(a), where host sites have one, uniform hopping
amplitude $t_{0}$ and chemical potential $\mu_{0}$ and impurity
sites, being pairing centers, have different hopping $t_{i}$ and
paring $\Delta_{i}$ amplitudes and chemical potentials $\mu_{i}$,
where $i=1,\dots,N$ labels impurities. Such a model is slightly more
general than the one that can be obtained from the spin-orbital model
describing $d^{4}$ host with $d^{2}$ impurities, because all the
impurities are equivalent, but one has to remember that orbital sector
is coupled with physical spins $S=1$. Hence if we are interested
in a purely orbital problem then, after averaging over non-uniform
spin configuration (if this is justified by weak entanglement of spins
and orbitals), the impurities can effectively differ from each other. 

The Hamiltonian of a 1D non-uniform superconductor, schematically
shown in Fig. \ref{fig:15}(a), was studied for possible topological
states \cite{Brz17R}. This was motivated by the fact that a homogeneous
Kitaev model \cite{Kit09} is topologically non-trivial as long as
$\mu_{0}<2t_{0}$. This non-triviality leads, in a system with open
edges, to so-called Majorana states localized at both edge of the
system. These states have zero energy so in a superconducting system
it is impossible to distinguish which one is an electron an which
one a hole state. They are interesting because they are topologically
protected and could be potentially used for creating qubits that are
robust against decoherence and thus for quantum computing. Signatures
of Majorana modes in real physical system were first reported in Ref.
\cite{Mou12}, where InSb nonowires contacted with $s$-wave superconductors
were experimentally studied. Therefore the question of possible topological
states in inhomogeneous Kitaev model is relevant since, firstly, physical
systems are often non-uniform, secondly, the spin-orbital system from
which this model originates could be a novel platform for obtaining
Majorana states realized by orbital degrees of freedom. From this
point of view finding analytical expressions to determine whether
a system of length $L$ with $N$ impurities is topologically non-trivial
is a relevant result of Ref. \cite{Brz17R}. Additionally, an important
simplification is by introducing variables that allow for strong reduction
of number of relevant parameters of the model and exhibiting hidden
Lorentz symmetry of topological phases, depicted in Fig. \ref{fig:15}(b).
It was thus possible to demonstrate that even a system with complete
disorder of impurities distribution has a non-vanishing area in parameter
space where a topological state is realized, what happens even for
small impurities concentration of the order of $2\%$. Consequently,
when such a system is opened the Majorana end-modes were observed
\cite{Brz17R}.

Non-uniform Kitaev model belongs to the same symmetry class as a uniform
one, i.e., it has a time-reversal ${\cal T}$ and particle-hole ${\cal C}$
symmetry and thus also a chiral symmetry ${\cal S}={\cal T}{\cal C}$.
From the general classification of topological states one gets that
such a model can have a non-trivial $\mathbb{Z}$ topological index,
in one dimension given by so called winding number. It can be defined
in a following way; in the eigenbasis of ${\cal S}$ operator Hamiltonian
${\cal H}_{k}$ of the model in momentum space (assuming translational
invariance with any unit cell) has a block-antidiagonal form with
two blocks given by matrices ${\bf u}_{k}$ and ${\bf u}_{k}^{\dagger}$.
A determinant of ${\bf u}_{k}$ matrix is a complex number and sets
a map from BZ, being a 1D sphere, to the complex plane. Such a map
can be non-trivial in the sense of homotopy groups, i.e., vector $\vec{v}_{k}$
given by real and imaginary part of determinant of ${\bf u}_{k}$
can rotate by a $n2\pi$ angle after one turn around the BZ, where
$n$ is an integer. It is not hard to guess that $n$ is equivalent
with the topological index $\mathbb{Z}$. In case of the considered
model the determinant is given by a formula, $\det{\bf u}_{k}\propto{\cal A}+{\cal B}\cos k+i{\cal C}\sin k$,
where ${\cal A}$, ${\cal B}$ and ${\cal C}$ are real constants
and $i$ is an imaginary unit. Thus, by changing $k$ from $0$ to
$2\pi$ we have three options; either vector $\vec{v}_{k}$ does not
make any rotation or it rotates once clockwise or anticlockwise. These
two last cases give a topological state and occur if only $|{\cal A}|<|{\cal B}|$
and ${\cal C}\not=0$. What is interesting, expressions for coefficients
${\cal A}$, ${\cal B}$ and ${\cal C}$ as functions of the parameters
$\{\mu_{0},t_{0}\}$ for the host and $\{\mu_{i},t_{i},\Delta_{i}\}_{i=1}^{N}$
for the impurities at positions $p_{i}$ in the unit cell of the length
$L$ can be obatained in an exact and closed form \cite{Brz17R}.

\begin{figure*}[t]
\begin{center}\includegraphics[width=1\textwidth]{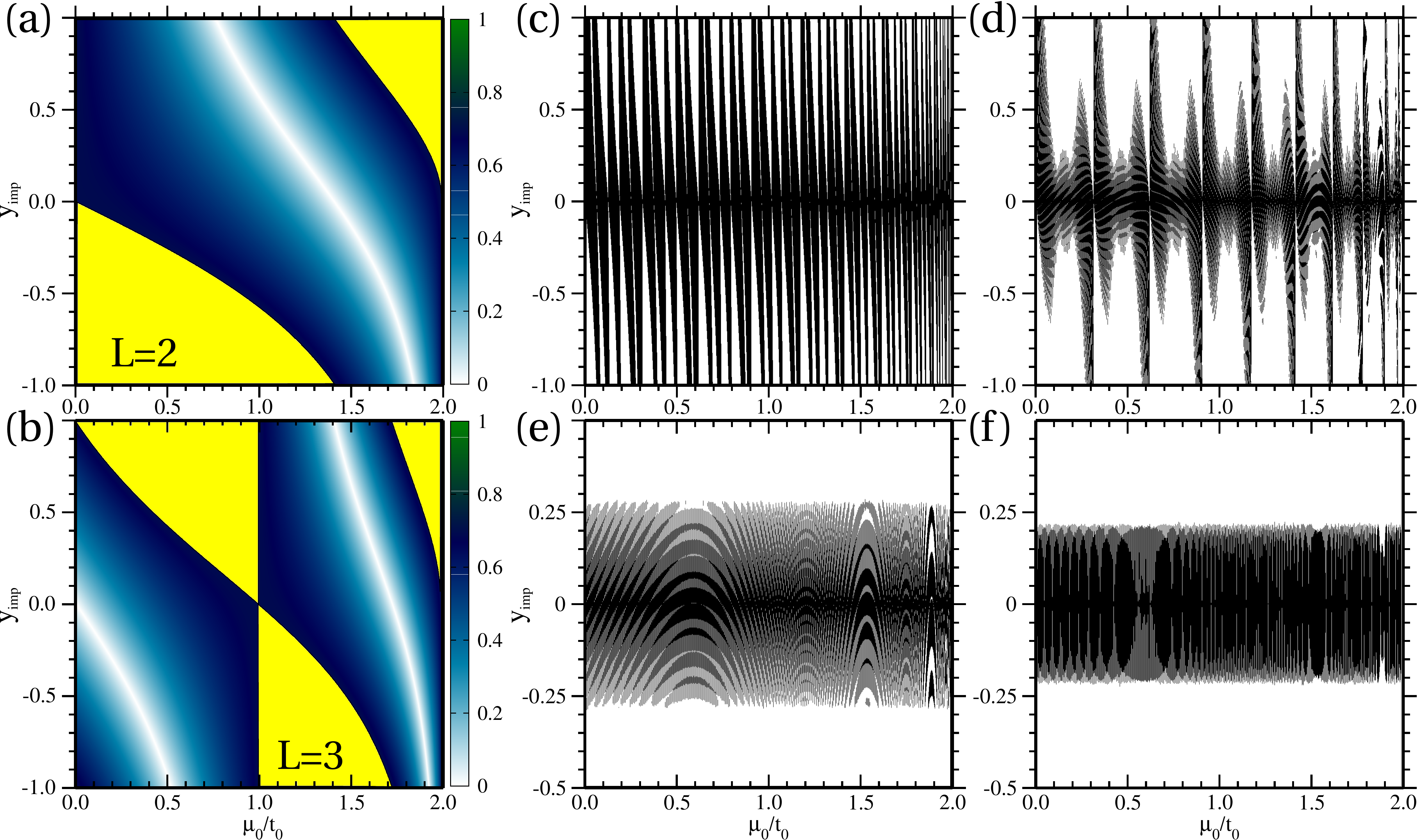}\end{center}\caption{Topologically stable domains (in blue with gradient or black), such
that $|{\cal A}|<1$, as functions of host parameter $\mu_{0}/t_{0}$
and a single reduced parameter of impurities $y_{i}\equiv y_{{\rm imp}}$.
Diagrams (a) and (b) \textendash{} high concentration of impurities,
single impurity in a unit cell of length $L=2,3$ \textendash{} blue
gradient indicates value of $|{\cal A}|$ according to the color scale,
$|{\cal A}|>1$ in the yellow area. Diagrams (c-f) \textendash{} low
concentration of impurities, doping $2\%$. (c) Ordered impurities
with two distances between NN impurities $d_{12}=40$ and $d_{21}=60$.
(d) System with binary disorder, for a given pair of neighboring impurities
distance $40$ or $60$ is picked random with the same probability.
(e) System with complete disorder of impurity distribution in a unit
cell of the size $L=1000$ and (f) the same system with random sign
of parameter $y_{i}=\pm y_{{\rm imp}}$. Results for systems with
disorder are averaged over $1000$ realizations of disorder. Plots
(c-f) consist only of black and white dots.\label{fig:16}}
\end{figure*}

Expressions for ${\cal A}$, ${\cal B}$ and ${\cal C}$ are simple
enough to present them in full glory. The biggest simplification is
achieved by a hyperbolic parametrization for impurities, i.e., for
any impurity $i$ we put $t_{i}=r_{i}\cosh\phi_{i}$ and $\Delta_{i}=r_{i}\sinh\phi_{i}$,
where by an analytic continuation this parametrization covers the
whole $t_{i}-\Delta_{i}$ planes excluding the diagonal $|t_{i}|=|\Delta_{i}|$,
being analogical to the light cones. With such parametrization we
get coefficients ${\cal B}=\cosh(2\sum_{i}\phi_{i})$ and ${\cal C}=\sinh(2\sum_{i}\phi_{i})$
as dependent only on the sum of angles $\phi_{i}$, whereas coefficient
${\cal A}$ depends on radii $r_{i}$, positions of impurities $p_{i}$
and host's parameters. These dependencies however can be reduced to
a dimensionless parameter $y_{i}$, related with every impurity $i$,
whose form is $y_{i}=t_{0}(\mu_{i}/r_{i}^{2}-\mu_{0}/t_{0}^{2})/\sqrt{4-\mu_{0}^{2}/t_{0}^{2}}$
and to one dimensionless parameter $\eta_{0}$ determining the host,
, $\eta_{0}=\arccos[\mu_{0}/(2t_{0})]$. Coefficient ${\cal A}$ can
be now written, using triangular matrices ${\bf M}_{1}$ and ${\bf M}_{2}$
and a diagonal matrix ${\bf Y}$, as ${\cal A}=\cos(L\eta_{0})+{\rm Tr}\{({\bf 1}-2{\bf M}_{1}{\bf Y})^{-1}{\bf M}_{2}^{T}{\bf Y}\}$,
where non-vanishing matrix entries are; $({\bf M}_{1})_{ij}=\sin(\eta_{0}d_{ij})$
and $({\bf M}_{2})_{ij}=\sin[\eta_{0}(L-d_{ij})]$ for $j\geqslant i$
and ${\bf Y}_{ii}=y_{i}$. Matrices ${\bf M}_{1,2}$ encode spatial
distribution of impurities by distances $d_{ij}\equiv p_{j}-p_{i}$
between impurity $i$ and $j$, so they contain some interference
of single-particle states localized between two impurities and $1/\eta_{0}$
can be treated as en effective Fermi length of the host.

As one can directly see from the form of coefficients ${\cal A}$,
${\cal B}$ and ${\cal C}$, a system that is in a topological state
can be modified in infinitely many ways and it will remain topological.
Especially, ${\cal B}$ and ${\cal C}$ depend only on the sum of
angles $\phi_{i}$, so one can freely modify $N-1$ angles provided
that one compensates these changes by the last angle. Therefore, we
have symmetries of the Lorentz type in every $t_{i}-\Delta_{i}$ plane
related with impurity $i$, which is depicted in Fig. \ref{fig:15}(b).
Discovery of such symmetries of topological phase is rather an unexpected
and non-trivial result for a system with disorder. Another non-trivial
symmetry is scaling of $\mu_{i}$ and $r_{i}^{2}$ at every impurity
by an arbitrary constant $\alpha_{i}$, which neither affects values
of $y_{i}$ parameters nor obviously angles~$\phi_{i}$.

Question about the topological state of the system comes down to question
about the values of ${\cal A}$, ${\cal B}$ and ${\cal C}$, which
still depend on many variables. Analysis of this problem can be simplified
if we notice that we always have ${\cal B}\geq1$. In such a case
if ${\cal A}<1$ and angles $\phi_{i}$ do not sum up to zero, then
the system is always topological. The area in the parameter space
where ${\cal A}<1$ we will call a topologically stable domain. Having
a general formula for ${\cal A}$ we can now study such domains for
different impurity configurations. Note that from the form of coefficient
${\cal A}$ it follows that for $y_{i}\equiv0$ we always have ${\cal A}\leq1$
independently on all other variables. Figures \ref{fig:16}(a-b) show
topological domains for a single impurity with a parameter $y_{1}=y_{{\rm imp}}$
for $L=2$ and $L=3$, so a high concentration of impurities. We see
that with increasing of the unit cell a topological domain gets fragmented
and these fragments always evolve around lines where ${\cal A}=0$.
Impassable boundary of any topological domain is always point $\mu_{0}=2t_{0}$,
being at the same time a boundary of topological phase for a uniform
Kitaev model.

Figures \ref{fig:16}(c-f) show the evolution of a topological domain
for $2\%$ concentration of impurities with increasing disorder of
their positions but with the same parameters $y_{i}\equiv y_{{\rm imp}}$
on each impurity, with exception of \ref{fig:16}(f), where the sign
of $y_{i}$ is random variable. Fig. \ref{fig:16}(c) concerns an
ordered system where the distances between neighboring impurities
alternate between two values $d_{12}=40$ and $d_{21}=60$, so it
is a dimerized system with two impurities in the unit cell. Topological
domain is strongly cut vertically into narrow legs stretching in wide
range of $y_{{\rm imp}}$. In Fig. \ref{fig:16}(d) we introduce a
binary Poisson disorder where distances between neighboring impurities
take random and equiprobable values $40$ or $60$, in such a way
that the total number of short and long distances is the same. Coefficient
${\cal A}$ is then averaged over many realizations of the disorder
and based on $\langle{\cal A}\rangle$ we determine the topological
domain. One can notice some similarities of this domain to an ordered
case with a difference that some of the legs are cut in vertical axis
and there appears a subtle interference pattern, which makes some
parts of this area full of holes, resembling Sierpi\'{n}ski carpet,
Cantor set, or other fractal structures. Increasing further the disorder
we obtain a domain shown in Fig. \ref{fig:16}(e), where impurities
are placed completely randomly, with a restriction that they are never
neighboring. As one can see, there are no vertical legs and the vertical
boundaries of the domain seem to be independent on $\mu_{0}$. Interference
patterns is clear and resembles many overlapping parabolas whose tips
seem to accumulate at few distinct values of $\mu_{0}$, which probably
is related with forming of charge density waves between the impurities
for chosen values of host's chemical potential. One can also notice
an asymmetry of the domain with respect to positive and negative $y_{{\rm imp}}$,
where this effect disappears when apart from position disorder we
randomize the sign of $y_{{\rm imp}}$ at each impurity. This case
is shown in Fig. \ref{fig:16}(f) where topological domain shrinks
in vertical direction and interference pattern has a form of long
and thin fingers of a trivial phase entering the domain. What is interesting,
the subtle character of topological domains \ref{fig:16}(e-f) does
not depend strongly on impurities concentration; only the width of
the domain grows with the decrease of their concentration as the number
of impurities $N$ is at the same time the maximal power of $y_{{\rm imp}}$
in the expression for coefficient~${\cal A}$.

Summarizing, the non-uniform Kitaev model showed interesting analytical
properties including hidden symmetries of the topological phase and
robustness against disorder. What more, it turned out that for any
disorder one can have topological states and consequently Majorana
states in open system if only the parameters of impurities $y_{i}$
are sufficiently close to zero. This is a condition of some kind of
resonance between impurities and the host that occurs when $t_{0}^{2}\mu_{i}\equiv(\Delta_{i}^{2}-t_{i}^{2})\mu_{0}$
and allows the system to igore the disorder inflicted by the impurities.

\section{Relationship of non-uniform Kitaev model with charge dilution\label{sec:Relationship-of-non-uniform}}

\begin{figure*}[t]
\begin{center}\includegraphics[width=1\textwidth]{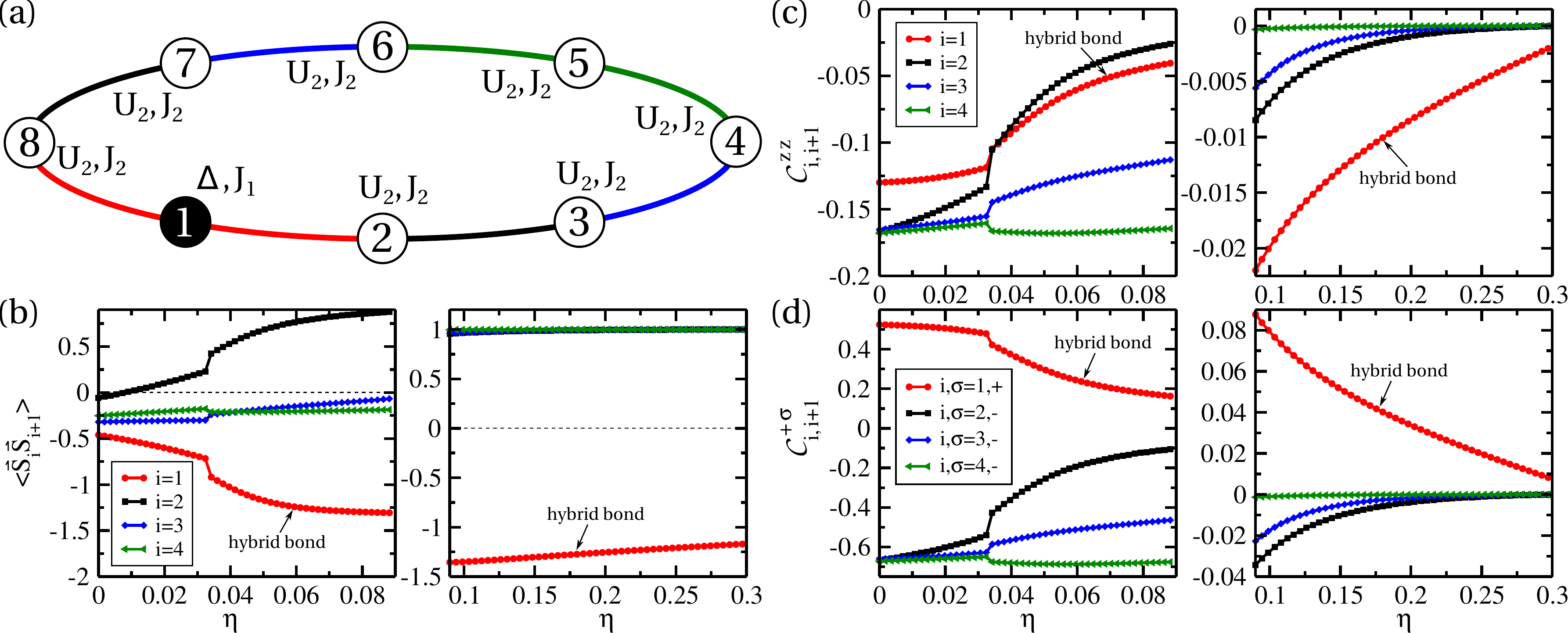}\end{center}\caption{(a) Schematic view of the system with $8$ sites with one $d^{2}$
impurity at site $i=1$ and seven $d^{4}$ host atoms at sites $i=2,\dots,8$.
Parameters of impurity are $\Delta$ and $J_{1}$ and of the host
$U_{2}$ and $J_{2}$. Color convention for bonds $\langle i,i+1\rangle$
is kept in plots (b-d). (b) Ground-state spin correlations $\langle\vec{S}_{i}\vec{S}_{i+1}\rangle$
as functions of $\eta=J_{2}/U_{2}$ in the AF regime (left plot) and
the FM regime (right plot) of the host. (c) and (d) Analogical plots
for spin-orbital covariances ${\cal C}_{i,i+1}^{zz}$ and ${\cal C}_{i,i+1}^{+\sigma}$,
where $\sigma=-$ for host bonds and $\sigma=+$ for hybrid bonds
around the impurity. \label{fig:17}}
\end{figure*}

The model described above originates from a spin-orbital system with
charge dilution, i.e., $d^{4}$ system doped with $d^{2}$ metal.
The spin-orbital model is however richer than the Kitaev one because
orbitals can be entangled with spins. Therefore, one has verify if
this entanglement can be neglected and if the pure orbital model,
with such parameters as follow from the superexchange, can become
topological. To address this issue the full spin-orbital model was
studied for a single impurity and seven host atoms \cite{Brz17doi}.
Schematic view of such a setup is presented in Fig. \ref{fig:17}(a),
where $U_{2}$ and $J_{2}$ are the Hubbard and Hund's interactions
of the host, $J_{1}$ is a Hund's interaction at the impurity and
$\Delta$ is characteristic excitation energy for hybrid bonds between
impurity and host, analogical to the one which was introduced for
orbital dilution \cite{Brz17}. Main result here is demonstrating
that if $\eta=J_{2}/U_{2}$ is sufficiently large then spin order
of the host becomes FM and entanglement between spin and orbitals
is small. Thus, in this parameter range the spin interaction can be
substituted by their averages in the FM state and then using the results
of \cite{Brz17R} one can tell whether the pure orbital model is topological.

In Fig. \ref{fig:17}(b) the ground state NN spin correlation obtained
by exact diagonalization were shown. As one can see already for $\eta\simeq0.09$
the host becomes FM although hybrid bonds remain AF. One should then
check if such a bond does not generate high spin-orbital entanglement
in the system. In order to do it spin-orbital covariances on the bonds
${\cal C}_{i,i+1}^{zz}$ and ${\cal C}_{i,i+1}^{+\sigma}$ were calculated.
Their form is given by: ${\cal C}_{i,i+1}^{zz}=\langle\vec{S}_{i}\vec{S}_{i+1}\tau_{i}^{z}\tau_{i+1}^{z}\rangle-\langle\vec{S}_{i}\vec{S}_{i+1}\rangle\langle\tau_{i}^{z}\tau_{i+1}^{z}\rangle$
and ${\cal C}_{i,i+1}^{+\sigma}=\langle\vec{S}_{i}\vec{S}_{i+1}\tau_{i}^{+}\tau_{i+1}^{\sigma}\rangle-\langle\vec{S}_{i}\vec{S}_{i+1}\rangle\langle\tau_{i}^{+}\tau_{i+1}^{\sigma}\rangle+c.c.$,
where $\sigma=\pm$, which follows from the form of spin-orbital terms
in the Hamiltonian. Zero value of these covariances means basically
that the wave function of the system can be factorized into spin and
orbital part. The behavior of these quantities as functions of $\eta$
was shown in Figs. \ref{fig:17}(c) and \ref{fig:17}(d). It turns
out that non-vanishing off-diagonal covariances ${\cal C}_{i,i+1}^{+\sigma}$
for host bonds are only ones with $\sigma=-$, whereas for hybrid
bonds the ones with $\sigma=+$, so only such covariances were presented
in Fig. \ref{fig:17}(d). As one can see the AF region for low $\eta$
is characterized by relatively high covariances, so one cannot decouple
spin and orbitals. On the other hand, in the FM regime, the covariances
are order of magnitude smaller and, especially for host sites, they
vanish quickly with growing $\eta$. A slightly higher entanglement
remains on hybrid bonds but one can expect that this effect will be
weakened for smaller concentrations of impurities because FM state
of the host will be effectively suppress spin fluctuations.

After averaging over spin the pure orbital model was solved by a Hartree-Fock
approximation for terms $\tau_{i}^{z}\tau_{i+1}^{z}$ written with
Jordan-Wigner fermions. In this way the system was mapped onto inhomogeneous
Kitaev model and then its topological non-triviality was confirmed
using the results from Sec. \ref{sec:Topological-phases-in}. Hence,
\cite{Brz17doi} shows that the spin-orbital model with charge dilution
can have a topological state in some cases and consequently orbital
Majorana states at the edges.

\section{Summary\label{sec:Summary}}

We discussed non-trivial cases of spin, orbital and topological orders
in models describing complex and strongly correlated transition metal
oxides. Within a spin-orbital model for transition metals in the $d^{9}$
configuration, i.e., the Kugel-Khomskii model, the noncollinear magnetic
phases were found, whose stability does not rely on SOC but only on
strong orbital fluctuations and spin-orbital entanglement \cite{Brz12,Brz13}.
On the other hand, a rigorous topological order was uncoveded in one-dimensional
spin-orbital model, which follows from a non-trivial separation of
spin and orbitals in a periodic system \cite{Brz14}. Studies of the
same system with additional Heisenberg spin coupling showed the presence
of quantum phase transition involving spontaneous dimerization of
the system \cite{Brz15}. This dimerization occurs mainly in the spin
sector due to orbital fluctuations. Therefore, this is a mechanism
related to the one that gives noncollinear phases in the Kugel-Khomskii
model.

Another interesting case are models of $d^{4}$ transition metal oxides
doped with $d^{3}$ or $d^{2}$ metals. First of these cases we call
orbital dilution \cite{Brz15X,Brz16}, because the dopant is effectively
deprived of orbital degrees of freedom, and the second one is the
charge dilution \cite{Brz17}, because the dopant has orbital degrees
of freedom but realized by empty, not double occupancy of one of the
$t_{2g}$ orbitals. Orbital dilution was found to lead to strong modification
of the spin-orbital order of a $d^{4}$ host in different ranges of
microscopic parameters both in classical and quantum limit, including
spin-orbit coupling at host sites. This leads again to phases with
noncollinear magnetic order around the dopants and host's spin-orbital
order is modified even for small doping ratios. On the other hand,
in the case of charge dilution, it was shown that dopants are the
source of orbital pairing terms. It turned that they increase orbital
fluctuations and can drastically change the order of the host, similarly
as it happens for the orbital dilution case. Works \cite{Brz15X,Brz16,Brz17}
are one of few theoretical contributions to this very interesting
direction of research which are hybrid transition metal oxides. What
more, they are inspiration for further experimental studies of the
systems with orbital or charge dilution. We argue that short-range
order around impurities could be investigated by the excitation spectra
at the resonant edges of the substituting atoms. We expect that the
spin-orbital correlations could emerge in the integrated RIXS spectra
providing information of the impurity-host coupling and of the short-range
order around the impurity.

Another aspect of hybrid oxides and charge dilution is that in one
dimensions a $d^{4}$ system with a $d^{2}$ doping can be connected
with non-uniform Kitaev model having topologically non-trivial phases
\cite{Brz17R}, even in the case of complete disorder of the dopants
\cite{Brz17doi}. The mapping of one model onto the other is possible
only under condition that spin-orbital entanglement is small, which
happens in the ferromagnetic regime of the host. On the other hand,
results obtained for the Kitaev model are more general and show hidden
symmetries of the topological phase involving, inter alia, Lorentz
transformations in local space of impurities parameters. Therefore,
works \cite{Brz17R,Brz17doi} not only contribute to fundamental understanding
of properties of topological phases but also postulate existence of
orbital Majorana states in an insulating, strogly correlated spin-orbital
systems. An experimental realization and detection of such Majorana
modes remains an open question, although analogical realization of
Majorana bound-states in a spin model was suggested recently \cite{Hof18}.
The detection of such states could by through spin transmission through
the magnetic region of a characteristic resonant length. Therefore,
the orbital Majorana states could be detected in a similar manner.

A different connection of physics of topological states with transition
metal oxides and spin-orbital order can be made in the limit of itinerant
magnetism described by double-exchange models. The stability of zigzag
magnetic order in a bilayer $d^{4}$ system doped with $d^{3}$ ions,
so a case of orbital dilution, can be attributed to direction hopping
forced by the symmetries of the $t_{2g}$ orbitals \cite{Brz15L}.
The system was studied using a double exchange model, where electrons
with one orbital flavor are subject to localization and interact,
via Hund's interaction, with electrons on the other orbitals which
remain itinerant \cite{Brz15L}. It was shown that different types
of zigzag phases are stable due to the process of formation of orbital
molecules which for some electron densities allow to reduce the kinetic
energy in the system. This happens in the region of the parameter
space where the uniform ferromagnetic phase competes with the antiferromagnetic
one, so the kinetic energy of itinerant electrons is comparable with
magnetic exchange of localized spins. On the other hand, one can focus
on the topological aspect of the itinerant electrons that moving among
localized spins ordered in a zigzag fashion interact with them via
Hund's exchange and spin-orbit coupling \cite{Brz17B}. It turns out
that by tuning these two parameters we can obtain wide regions of
metallic, insulating an semi-metallic phases, where these last ones
are topologically non-trivial. Their non-triviality follows from the
presence of topologically protected Dirac points. These can be points
with higher than a double degeneracy and the topological protection
can arise from more than one topological charge present at the Dirac
point. The source of exotic features of topological semi-metallic
phases is the nonsymmorphic character of the zigzag magnetic order
\textendash{} it has a symmetry of a mirror reflection with a shift
of half of a lattice vector. What more, this symmetry has a tendency
to glue Dirac points together and form multiple degenerate point which
engage another symmetry, being however non-unitary (nor antiunitary),
acting on the level of determinant of the Hamiltonian. Therefore,
some new fundamental understanding of topological phases with nonsymmorphic
symmetries was developed in Ref. \cite{Brz17B}. It can be an inspiration
for experimental groups searching for materials which are both topological
and magnetic, for instance antiferromagnetic topological insulator
whose discovery was reported recently \cite{Otr2018}. Another interesting
option is interfacing such a planar magnetic system with a superconductor
to induce superconductivity in presence of nonsymmorphic symmetries,
as discussed in Ref. \cite{Brz18-nodal}. This can lead to an exotic
nodal-loop superconductivity even in proximity of a simple $s$-wave
superconductor. This shows that interfacing magnetic or superconducting
order with topology can lead to unexpected and interesting phases
of matter. 

\ack{}{}

I thank Andrzej M. Ole\'s, Mario Cuoco and Jacek Dziarmaga for friendly
and fruitful collaboration. This work is supported by the Foundation
for Polish Science through the IRA Programme co-financed by EU within
SG OP Programme. I acknowledge support by Narodowe Centrum Nauki (NCN,
National Science Centre, Poland) Project No. 2016/23/B/ST3/00839 and
by the European Union\textquoteright s Horizon 2020 research and innovation
program under the Marie Sklodowska-Curie Grant Agreement No. 655515,
project UFOX: Unveiling complexity in Functional hybrid Oxides.

\providecommand{\newblock}{}

\end{document}